\documentclass[prd,aps,a4paper,nofootinbib,twocolumn,showpacs,showkeys]{revtex4}  

\newif\ifusesec
\usesectrue  
   
\usepackage{graphicx} 
\usepackage{mathrsfs}
\usepackage{latexsym,amsmath,amsfonts,amssymb}
\usepackage{multirow}

\newcommand{\beq}{\begin{equation}}
\newcommand{\eeq}{\end{equation}}
\def\rightcontract{\mathop{\hbox{\vrule width0.5pt height6pt
  \vrule height0.5pt width6pt}}}

\def\R{\mathbb{R}}
\def\T{\mathbb{T}}
\begin{document}

\title{Gyroscope precession along general timelike geodesics in a Kerr black hole spacetime}

\author{Donato \surname{Bini}$^{1,2}$}
\author{Andrea \surname{Geralico}$^{1,2}$}
\author{Robert T. \surname{Jantzen}$^{2,3}$}

\affiliation{
$^1$Istituto per le Applicazioni del Calcolo ``M. Picone'', CNR, I-00185 Rome, Italy\\
$^2$ICRANet, Piazza della Repubblica 10, I-65122 Pescara, Italy \\
$^3$Department of Mathematics and Statistics, Villanova University, Villanova, PA 19085, USA
}

\date{\today}

\begin{abstract}
The precession angular velocity of a gyroscope moving along a general geodesic in the Kerr spacetime is analyzed using the geometric properties of the spacetime.
Natural frames along the gyroscope world line are explicitly constructed by boosting frames adapted to fundamental observers. 
A novel geometrical description is given to Marck's construction of a parallel propagated orthonormal frame along a general geodesic, identifying and clarifying the special role played by the Carter family of observers in this general context, thus extending previous discussion for the equatorial plane case.
\end{abstract}


\maketitle

\section{Introduction}

The recent discovery of gravitational waves by LIGO \cite{ligo,Abbott:2016blz} has emphasized that the most promising sources of gravitational radiation are coalescing binary systems made of spinning compact objects during the whole process (inspiral, merger, and ringdown).   
As a consequence, the relativistic community has experienced a renewed interest in all those rotational effects associated with the gravitational interaction between two such bodies, namely orbital, spin-orbital, and spin-spin effects, in order to build more and more accurate templates for gravitational wave emission profiles.
Spin couplings have been computed (only) by using standard approximation schemes, like post-Newtonian (PN) theory through a certain PN order both in the framework of Hamiltonian dynamics \cite{Damour:2007nc,Barausse:2009aa}, perturbation theory \cite{Dolan:2013roa,Bini:2014ica,Bini:2015mza,Akcay:2016dku} (only for motion along circular and eccentric equatorial orbits around a non-spinning black hole), and by using effective field theory techniques \cite{Levi:2014sba,Levi:2015msa,Levi:2016ofk}.
In all cases a key role is played by the spin precession angular velocity $\Omega_{\rm(prec)}$ of one spinning body with spin vector $S$ in the gravitational field of its companion. 
For instance, in the Hamiltonian description this coupling is taken into account by the spin-orbit Hamiltonian $H_{SO}=\Omega_{\rm(prec)}\cdot S$.
The problem of determining $H_{SO}$ as a function of the coordinates, conjugate momenta and spin has been successfully addressed for ADM coordinates in Ref.~\cite{Damour:2007nc}, followed by Ref.~\cite{Barausse:2009aa}, where the constrained Hamiltonian of a spinning test particle was derived to linear order in the particle's spin by using standard Boyer-Lindquist coordinates.

A general relativistic model describing the interaction of a small (test) spinning body with a spinning black hole is provided by the Mathisson-Papapetrou-Dixon (MPD) model \cite{Mathisson:1937zz,Papapetrou:1951pa,Dixon:1970zza}, which in the pole-dipole approximation accounts for spin couplings only at linear order in the particle's spin. An iterative prescription, implicit in the MPD construction, then allows the incorporation of the quadrupole moment (and higher multipole moments) as well as taking into account affects which are higher order in spin.
According to such a model, in the pole-dipole approximation the orbit of a spinning object deviates from geodesic motion due to acceleration effects arising from spin-curvature couplings, the spin vector being parallel transported along the path.
Therefore, the evolution of the spin vector to first order in spin is that of a test gyroscope moving along a geodesic of the background spacetime, and $\Omega_{\rm(prec)}$ is the precession angular velocity measured with respect to a Cartesian-like frame (defined all along its orbit) whose axes are aligned with the ``fixed stars" at spatial infinity.
In this work we will study the precession of a test gyroscope orbiting a Kerr black hole in generic geodesic motion, completing a program developed in a series of previous papers \cite{Bini:2016iym,Bini:2016ovy}.

To make a long story short and give a more precise context to our research we recall that gyroscope precession has attracted attention since the pioneering works of Lense and Thirring \cite{thirring1,thirring2,lense-thirring} and Schiff \cite{schiff}, leading to the Gravity Probe B mission \cite{gpb1,gpb2} which launched a ``test gyroscope'' moving along a geodesic with its spin vector parallely propagated along it.
Most theoretical investigations of this phenomenon have been limited to gyroscopes moving along circular equatorial plane orbits \cite{RindlerPerlick:1990,Iyer:1993qa,Jantzen:1992rg,Bini:1994,Bini:1997ea,Bini:1997eb,Bini:2002mh} due to their extremely simple geometrical properties.
More recently, the generalization of previous results to both bound and unbound equatorial plane eccentric geodesic orbits in a Kerr spacetime has been analyzed in Refs.~\cite{Bini:2016iym,Bini:2016ovy}. 
For gyroscopes in generic orbits around a Kerr black hole, very little work has been done but 
Ref.~\cite{Ruangsri:2015cvg} deserves mention. This (mostly numerical) study is performed in the frequency domain using functional techniques developed in Ref.~\cite{Drasco:2003ky} for bound geodesics, which are characterized by three fundamental frequencies \cite{Schmidt:2002qk,Fujita:2009bp}.

The main difficulty in extending the results for planar motion is that one must deal with generic rotations instead of the simpler planar rotations, complicating all the calculations. 
For equatorial plane orbits where the spin precession reduces to an azimuthal rotation characterized by a single scalar angular velocity directly interpretable as a frequency, 
this was accomplished simply by using the readily available frame found by Marck \cite{marck1,marck2}, which in that case is rigidly attached to the natural static spherical frame modulo a pair of boosts in the radial and azimuthal directions, taking advantage of a parallel transported direction associated with the Killing-Yano 2-form which exists for the Kerr spacetime.
Instead, for the general nonequatorial plane orbits considered in the present paper,
the angular velocity vector describing the spin precession has three nontrivial components, describing the time rate of change of the composition of the rotations needed to map from ``body-fixed'' axes in the gyroscope local rest frame to the axes in the rest frame of the  ``fixed stars'', modulo the unique boost between these two frames. 
One has the choice of continuing to use Marck's frame as an intermediary in evaluating the angular velocity of a parallel transported spin vector relative to boosted static spherical axes, or of simply directly evaluating the angular velocity of those boosted axes. We take both approaches and compare their relationship.

As with many questions in general relativity, where everything is ``relative", the definition of spin precession is a relative notion---one must compare the direction of the parallel transported spin vector of a gyroscope in geodesic motion in the Kerr spacetime in its local rest frame with a set of axes which are locked onto the ``distant stars" by the stationary symmetry, in a different local rest frame associated with static observers  in relative motion. By boosting the static observer axes to the gyroscope local rest frame by the unique boost associated with the relative motion, one removes the effects of stellar aberration due to that motion, but one must also take into account the rotation of the static spherical axes which occurs as the gyroscope moves in the angular direction with respect to the static grid associated with Boyer-Lindquist spherical-like coordinates. This is accomplished by introducing the naive Cartesian frame associated with the spherical frame as in flat space in order to lock the static axes to the distant sky at spatial infinity as one moves around in the spacetime. Combining these two effects leads to a way to quantify the precession of the gyroscope axes with respect to the distant sky in its local motion. 

After reviewing the description of timelike geodesics in Boyer-Lindquist coordinates, we introduce three fundamental observer families in the Kerr spacetime and their associated adapted frames: static observers, zero angular momentum observers (ZAMOs) and Carter observers. These are all related to each other by boosts in the azimuthal angular direction associated with the rotational Killing vector field. Modulo these boosts, they all share the same orthonormal spatial frame  obtained by normalizing the Boyer-Lindquist spherical-like coordinate frame vectors. This frame has its axes locked to the distant sky at spatial infinity, but only the static observers following the timelike Killing vector world lines see an unchanging distant sky from the null geodesics arriving from spatial infinity. Using the Cartesian-like frame corrects this orbital contribution to the rotation of the spherical axes due to motion in the coordinate grid.

Next, we compute the gyroscope precession angular velocity as measured by each family of observers in their natural frames, and determine their relationships.  The static observers are locked to the ``fixed stars" and as such are key to measuring precession with respect to them, so they are the preferred family of ``fiducial observers,'' but for motion relative to those observers, one needs a Cartesian-like frame rather than a spherical frame in order to lock the reference directions with respect to those fixed stars as the angular location changes.

We then discuss several examples of both general bound and unbound off-equatorial geodesic orbits as well as the special cases of orbits at constant radius (spherical and polar) and those constant polar angle $\theta$.
Subsequently, we will explicitly find the relation between the boosted static frame and Marck's parallely propagated frame.
The latter reduces the equations of parallel transport to a single ordinary differential equation involving a scalar angular velocity which may be expressed in terms of the constants of the motion characterizing the geodesic.
Finally, Appendix A shows how the kinematical properties of both static observers and ZAMOs affect precession, while Appendix B demonstrates that for general motion Marck's frame loses its special property of diagonalizing the curvature tidal matrices which holds in the equatorial plane orbit case.

We use standard notation with $G=1=c$, Greek indices running from $0$ to $3$, while Latin indices run from $1$ to $3$. The signature of the spacetime metric is $- + + +$.

\section{General geodesic motion} 

Following the notation of Misner, Thorne and Wheeler \cite{Misner:1974qy},
the Kerr metric written in Boyer-Lindquist coordinates $(t,r,\theta,\phi)$ is given by
\begin{eqnarray}
\label{K1}
ds^2&=&g_{\alpha\beta}dx^\alpha dx^\beta\nonumber\\
&=&-dt^2+\frac{\Sigma}{\Delta}dr^2+\Sigma\, d\theta^2 +(r^2+a^2)\sin^2\theta\, d\phi^2\nonumber\\
&& +\frac{2Mr}{\Sigma}(dt-a\sin^2\theta\, d\phi)^2\,,
\end{eqnarray}
where $M$ and $a$ are the mass and the specific angular momentum of the  source, respectively, and
\beq\label{K2}
\Sigma=r^2+a^2\cos^2\theta\,,\qquad \Delta=r^2-2Mr+a^2\,.
\eeq
Rotational properties of this spacetime are associated with the azimuthal coordinate $\phi$ associated with the axisymmetric Killing vector field $\partial_\phi$.
The outer horizon occurs at the zero $r_+=M+\sqrt{M^2-a^2}$ of $\Delta$, while the ergosphere occurs at at the zero $r_E=M+\sqrt{M^2-a^2\cos^2\theta}$ of $g_{tt}$.

\subsection{Timelike geodesics}

A  geodesic timelike world line has a 4-velocity unit tangent vector $U=U^\alpha \partial_\alpha$ with coordinate components $U^\alpha=dx^\alpha/d\tau$ which  can be expressed using the Killing symmetries \cite{Carter:1968ks,Chandrasekhar:1985kt} 
 as a system of first order differential equations
\begin{eqnarray}
\label{geo_eqs}
\frac{d t}{d \tau}&=& \frac{1}{\Sigma}\left[aB+\frac{(r^2+a^2)}{\Delta}P\right]\,,\nonumber \\
\frac{d r}{d \tau}&=&\epsilon_r \frac{1}{\Sigma}\sqrt{R}\,,\nonumber \\
\frac{d \theta}{d \tau}&=&\epsilon_\theta \frac{1}{\Sigma}\sqrt{\Theta}\,,\nonumber \\
\frac{d \phi}{d \tau}&=& \frac{1}{\Sigma}\left[\frac{B}{\sin^2\theta}+\frac{a}{\Delta}P\right]\,,
\end{eqnarray}
where $\tau$ is a proper time parameter along the geodesic,
$\epsilon_r$ and $\epsilon_\theta$ are sign indicators, and
\begin{eqnarray}
\label{geodefs}
P&=& E(r^2+a^2)-L\,a\,,\nonumber\\
B&=& L-aE \sin^2\theta\,, \nonumber\\
R&=& P^2-\Delta (r^2+K)\,,\nonumber\\
\Theta&=&K-a^2\cos^2\theta-\frac{B^2}{\sin^2\theta}\,.
\end{eqnarray}
Here $E$ and $L$ denote the conserved Killing energy and angular momentum per unit mass and $K$ is a separation constant, usually called the Carter constant.
In place of $K$ one often uses
\beq
Q= K - (L-aE)^2\equiv K-x^2\,,
\eeq
which vanishes for equatorial plane orbits.
Corresponding to the 4-velocity vector field $U$
is the index-lowered 1-form
\begin{eqnarray}
\label{Ugeogen}
U^\flat &=& -E\, dt +\frac{\Sigma}{\Delta}  \dot r \,dr +\Sigma \dot \theta \,d\theta +L \,d\phi\,.
\end{eqnarray}
 In what follows we will use the overdot notation $\dot f=df/d\tau$ for the proper time derivative along the geodesic.

In addition to the generic case of nonequatorial plane motion, 
we will also consider some special classes of such orbits which have interesting behavior, namely orbits at constant radius $r=r_0$, $\dot r=0$ (spherical orbits, generalizing the circular orbits in the equatorial plane) and orbits at constant polar angle $\theta=\theta_0$, $\dot \theta=0$. The former include the interesting physical case of precessing polar orbits which model the GPB experiment.

\subsubsection{Spherical orbits}

Examining the geodesic equations \eqref{geo_eqs}, spherical geodesics are characterized by constant $r=r_0$ and $R=0=dR/dr$, so that \cite{Wilkins:1972rs}
\begin{eqnarray}
E&=&\frac{(r-M)(r^2+K)+r\Delta}{2r\sqrt{\Delta(r^2+K)}}\,, \nonumber\\
L&=&\frac{1}{2ar\sqrt{\Delta(r^2+K)}}
\{r(r^2+a^2)\Delta \nonumber\\
&&\qquad +(r^2+K)[M(r^2-a^2)-r\Delta]\}\,,
\end{eqnarray}
with arbitrary $K\geq x^2$ (or $Q\geq0$), the equality corresponding to circular equatorial orbits, i.e., $K_-\leq K\leq K_+$ with 
\beq
K_\pm=r\,\frac{Mr^2(r-3M)+ra^2(r+M)\pm2a\Delta\sqrt{Mr}}{r(r-3M)^2-4Ma^2}\,.
\eeq
The motion thus oscillates between the (supplementary) values of $\theta$ which satisfy the equation $\Theta=0$, crossing repeatedly the equatorial plane. In terms of the variable $z=\cos^2\theta$ the previous condition can be rewritten as
\beq
a^2(1-E^2)z^2-[a^2(1-E^2)+J^2]z+Q=0\,,
\eeq
with solutions
\beq
z_\pm=\frac{a^2(1-E^2)+J^2\pm\sqrt{D}}{2a^4(1-E^2)^2}\,,
\eeq
where 
$D=[a^2(1-E^2)+J^2]^2-4a^4(1-E^2)^2Q$,
$J=\sqrt{L^2+Q}$ denotes the total angular momentum per unit mass of the particle, such that $L=J\cos\iota$ and $\sqrt{Q}=J\sin\iota$ in terms of the ``inclination'' angle $\iota$. 
The range of allowed values for $z$ is then between zero and the smaller root $z_-$, implying that $\theta_-\leq \theta\leq \pi-\theta_-$.

Polar orbits are the subclass which intersect the symmetry axis of the black hole, where $\sin \theta=0$ \cite{stog-tsou}. From Eqs.~\eqref{geo_eqs} this requires zero angular momentum $L=0$. As a consequence, as shown in Ref.~\cite{stog-tsou},  the associated orbits have only $r$ and $\theta$ motions  with respect to ZAMOs (i.e., they co-rotate azimuthally with respect to the ZAMOs).
Spherical polar orbits are thus characterized by
\begin{eqnarray}
E&=&\Delta\left[\frac{r}{(r^2+a^2)[r\Delta-M(r^2-a^2)] }\right]^{1/2}\,, \nonumber\\
K&=&r\frac{Mr(r^2-a^2)+a^2\Delta}{r\Delta-M(r^2-a^2)}\,.
\end{eqnarray}

\subsubsection{Orbits at constant $\theta=\theta_0$}

Orbits with constant $\theta=\theta_0$ requiring $\Theta=0=d\Theta/d\theta$ are characterized by
\begin{eqnarray}
E&=&\frac{K+a^2(1-2\cos^2\theta)}{2a\sin\theta\sqrt{K-a^2\cos^2\theta}}\,, \nonumber\\
L&=&-\frac{(K-a^2)\sin\theta}{2\sqrt{K-a^2\cos^2\theta}}\,,
\end{eqnarray}
with arbitrary $K>a^2\cos^2\theta$.
The special value $K=a^2$ gives $E=1$ and $L=0$.

\section{Fundamental observer families and adapted frames}

Any test family of observers are characterized by a unit timelike 4-velocity field $u$ whose integral curves are the world lines of those observers, and whose local rest spaces will be designated by $LRS_u$. A stationary axisymmetric spacetime has several natural observer families associated with its special geometry.
In the Kerr case three families of observers play a key role, and are easily described in terms of the Boyer-Lindquist coordinates because they are adapted to the Killing symmetries. The static observers follow the integral curves of the (stationary) Killing vector field $\partial_t$, namely the Boyer-Lindquist time lines, while the world lines of the zero angular momentum observers (ZAMOs) are orthogonal to the time coordinate hypersurfaces. Finally the Carter observers are key to the separability of the geodesic equations which allow their exact solution, as well as being fundamentally important to the algebraic properties of the curvature tensor. All three families differ only by relative azimuthal motion, and hence their natural adapted frames are all related by relative boosts in the $t$-$\phi$ plane of the tangent space.

The static observers, 
which exist only in the spacetime region outside the black hole ergosphere where $g_{tt}<0$,  form a congruence of accelerated, nonexpanding and locally rotating world lines. 
They are, however, nonrotating with respect to observers at rest at spatial infinity and have 4-velocity $u=m$ where
\beq
m=\frac{1}{\sqrt{-g_{tt}}}\,\partial_t
=\left(1-\frac{2Mr}{\Sigma}\right)^{-1/2}\,\partial_t\,.
\eeq
An orthonormal frame adapted to $m$ is
\begin{eqnarray}
\label{static_triad}
e(m)_1 &=&\frac{1}{\sqrt{g_{rr}}}\,\partial_r =\sqrt{\frac{\Delta}{\Sigma}} \,\partial_r
\equiv e_{\hat r}
\,,\nonumber\\ 
e(m)_2 &=&\frac{1}{\sqrt{g_{\theta\theta}}}\,\partial_\theta = \frac{1}{\sqrt{\Sigma}}\,\partial_\theta
\equiv e_{\hat \theta}
\,,\nonumber\\
e(m)_3 &=&\frac{1}{\sqrt{g_{\phi\phi}-{g_{t\phi}{}^2}/{g_{tt}}}}
    \left(\partial_\phi-\frac{g_{t\phi}}{g_{tt}}\partial_t\right) \nonumber\\
&=& \frac{\sqrt{\Delta -a^2 \sin^2\theta}}{\sin \theta \sqrt{\Delta \Sigma}} \left(\partial_\phi-\frac{2Mar\sin^2\theta}{\Delta -a^2 \sin^2\theta} \partial_t  \right)
\,. \nonumber\\
\end{eqnarray}
The relative decomposition of the geodesic 4-velocity $U$ with respect to the static observers is
\beq
\label{Uthd}
U=\gamma(U,m)\left[m+\nu(U,m)^{a}  e_a(m)\right]\,,
\eeq
with
\begin{eqnarray}
\label{Uthd2}
&&[\gamma(U,m) \nu(U,m)^{a}]=\nonumber\\
&&
\left[
\sqrt{\frac{\Sigma}{\Delta}}\dot r, \sqrt{\Sigma}\, \dot \theta ,
\frac{1}{\sqrt{\Sigma\Delta}}\left(\frac{L\sqrt{\Sigma-2Mr}}{\sin\theta}+\frac{2aMrE\sin\theta}{\sqrt{\Sigma-2Mr}}\right)
\right]\nonumber\\
\end{eqnarray}
and
\beq
\gamma(U,m)=\frac{E\sqrt{\Sigma}}{\sqrt{\Sigma-2Mr}}\,.
\eeq

The ZAMOs are instead locally nonrotating (but locally rotating in the azimuthal direction in the same sense as the rotation of the black hole) and exist everywhere outside of the outer horizon,
They have 4-velocity $u=n$ where
\begin{eqnarray}\label{ncoord}
n&=&\sqrt{-g^{tt}}\,\left(\partial_t+\frac{g^{t\phi}}{g^{tt}}\partial_\phi\right)
\nonumber\\
&=& \sqrt{\frac{A}{\Delta\Sigma}}\,\left(\partial_t+\frac{2aMr}{A}\partial_\phi\right)
\,,
\end{eqnarray}
where 
\beq
A=(r^2+a^2)^2-a^2\Delta\sin^2\theta\,.
\eeq
The normalized spatial coordinate frame vectors 
\begin{eqnarray}\label{zamotriad}
e(n)_1&=&e_{\hat r}\,, \qquad
e(n)_2=e_{\hat \theta}\,, \nonumber\\
e(n)_3&=&\frac{1}{\sqrt{g_{\phi\phi}}}\,\partial_\phi 
=  \frac{\sqrt{\Sigma}}{\sin\theta \sqrt{A}} \,\partial_\phi
\equiv e_{\hat \phi}
\end{eqnarray}
together with $n$ form an orthonormal adapted frame.
A boost along $e_{\hat \phi}$ maps $n$ into $m$, i.e., 
\beq
m=\gamma(m,n)[n+\nu(m,n)]\,,
\eeq
with relative velocity in the opposite azimuthal direction as the rotation of the black hole associated with the sign of $a$ (resisting the ``dragging of inertial frames")
\beq
\nu(m,n)=-\frac{2Mr}{\Sigma} \left(\frac{a \sin\theta }{\sqrt{\Delta}}\right)\,e_{\hat\phi}\,,
\eeq
and associated Lorentz factor $\gamma(m,n)$,
so that ZAMOs and static observers share the same $r$-$\theta$ 2-plane of their local rest spaces.
The relative decomposition of the geodesic 4-velocity $U$ with respect to the ZAMOs is
\beq
\label{Uzamo}
U=\gamma(U,n)\left[n+\nu(U,n)^{a}  e_a(n)\right]\,,
\eeq
with
\beq
[\gamma(U,n) \nu(U,n)^{a}]
=\left[
\sqrt{\frac{\Sigma}{\Delta}}\dot r, \sqrt{\Sigma}\, \dot \theta , \frac{L\sqrt{\Sigma}}{\sin\theta\sqrt{A}}\right]
\eeq
and
\beq
\gamma(U,n)=\frac{1}{\sqrt{\Sigma\Delta}}\left[\sqrt{A}E-\frac{2aMrL}{\sqrt{A}}\right]\,.
\eeq
The kinematical properties of both the static observers and ZAMOs are briefly reviewed in Appendix A.

The third useful observer family in the Kerr spacetime,  the Carter family of observers $u=u_{\rm (car)}$,  is intimately connected with its geometry  (separability of the geodesic equations).
They are boosted in the opposite azimuthal direction from the static observers compared to the ZAMOs in order to ``comove" with the black hole, their angular velocity at the outer horizon being defined as that of the black hole itself. Their
$4$-velocity $u_{\rm (car)}$ lies in the intersection of two special 2-planes: the plane $T_2$ spanned by the temporal and azimuthal Killing vectors ${\rm span}\{ \partial_t , \partial_\phi \}={\rm span}\{ u_{\rm (car)},\partial_\phi\}$ and the plane $N_2$ spanned by the (two, repeated) principal null directions of the spacetime 
${\rm span}\{l,k\}={\rm span}\{u_{\rm (car)},\partial_r\}$. 
Its coordinate components may be read off from the expressions
\begin{eqnarray}
\label{car_obs}
u_{\rm (car)} &=&\frac{r^2+a^2}{\sqrt{\Delta \Sigma}}\left(\partial_t +\frac{a}{r^2+a^2}\,\partial_\phi  \right)\,,\nonumber\\
u_{\rm (car)}^\flat &=& -\sqrt{\frac{\Delta}{\Sigma}}(dt -a \sin^2 \theta \,d\phi)\,.
\end{eqnarray}
Decomposing it with respect to the static observers 
\beq\label{carsta}
u_{\rm (car)}= \gamma(u_{\rm (car)},m) [m+\nu(u_{\rm (car)},m)]\,,
\eeq
leads to the relative velocity 
\beq
\nu(u_{\rm (car)},m) = \frac{a\sin \theta}{\sqrt{\Delta}} e(m)_3\,.
\eeq
Comparing \eqref{ncoord} and \eqref{car_obs} shows that $u_{\rm(car)}$ lies between $n$ and $m$ as claimed above.

The orthogonal (azimuthal) direction to $u_{\rm (car)}$ within the 2-plane $T_2$ has coordinate components
\begin{eqnarray}
\bar u_{\rm (car)} &=& \frac{a\sin \theta}{\sqrt{\Sigma}}\left(\partial_t +\frac{1}{a\sin^2\theta}\,\partial_\phi  \right)\,,\nonumber\\
\bar u_{\rm (car)}^\flat  &=& -\frac{a\sin\theta}{\sqrt{\Sigma}}\left(dt -\frac{r^2+a^2}{a} \,d\phi  \right)\,.
\end{eqnarray}
A spherical orthonormal frame adapted to $u_{\rm (car)}$ is obtained by using the triad boosted from the either the ZAMO or static observer spherical frame along the azimuthal direction
\begin{eqnarray}
E_1(u_{\rm (car)}) &=& e_{\hat r}\,,\qquad 
E_2(u_{\rm (car)}) =e_{\hat \theta}\,,\nonumber\\
E_3(u_{\rm (car)}) &=& \bar u_{\rm (car)}\,. 
\end{eqnarray}
The relative decomposition of the geodesic 4-velocity $U$ with respect to the Carter observers is
\beq
\label{Uucar}
U=\gamma(U,u_{\rm (car)})\left[u_{\rm (car)}+\nu(U,u_{\rm (car)})^{a}  E_a(u_{\rm (car)})\right]\,,
\eeq
with
\beq
\label{Uucar1}
[\gamma(U,u_{\rm (car)}) \nu(U,u_{\rm (car)})^{a}]
=\left[
\sqrt{\frac{\Sigma}{\Delta}}\dot r, \sqrt{\Sigma}\, \dot \theta , \frac{B}{\sqrt{\Sigma}\sin \theta}
\right]
\eeq
and
\beq
\gamma(U,u_{\rm (car)})=\frac{P}{\sqrt{\Delta \Sigma}}\,.
\eeq

\section{Adapted frames and gyroscope precession}

Given a family of observers with 4-velocity $u$ and adapted orthonormal frame $\{u,e(u)_a\}$ one can form an adapted frame along a general timelike geodesic with 4-velocity $U$ by boosting the vectors $e(u)_a$ onto the local rest space of $LRS_U$.
Following the notation of \cite{Jantzen:1992rg}, a relative observer boost $B(U,u)$ mapping $LRS_u$ to $LRS_U$ is described by
\begin{eqnarray}\label{Blrs}
U&=&\gamma (U,u)[u+\nu(U,u)^a e(u)_a]\nonumber\\
E(U,u)_a&=& e(u)_a +\frac{U+u}{1+\gamma (U,u)} (U\cdot e(u)_a)\nonumber\\
&\equiv&B_{\rm (lrs)}(U,u)e(u)_a\,.
\end{eqnarray}
Letting $u=m$ and $u=n$ respectively one obtains the triads 
$E(U,m)_a=B_{\rm(lrs)}(U,m) e(m)_a$ and $E(U,n)_a=B_{\rm(lrs)}(U,n) e(n)_a$,
the two being related by a spatial rotation in $LRS_U$ due to the relative azimuthal rotation between their two 4-velocities which leads to a Wigner rotation $R(m,n)$ resulting from successive boosts along distinct directions compared to a direct boost \cite{Wigner:1939cj}
\begin{eqnarray}
E(U,m)_a &=& B(U,m) e(m)_a = B(U,m) B(m,n) e(n)_a\nonumber\\
 &=& B(U,m) B(m,n) B(U,n)^{-1} E(U,n)_a\nonumber\\
 &=& R(m,n) E(U,n)_a\,.
\end{eqnarray}

Our goal is to  study the evolution of the direction of the parallel transported spin vector $S\in LRS_U$ compared to axes which are fixed with respect to the ``fixed stars" modulo relative motion with respect to the static observers which see these stars as fixed in their local sky, requiring that stellar aberration must be removed. If $E(U)_a$ denotes a generic orthonormal frame in $LRS_U$, then 
\beq
S=S^a E(U)_a\,,
\eeq
and
\beq
\label{DUS}
\frac{DS}{d\tau}=\frac{dS^a}{d\tau}E(U)_a+S^a\frac{D}{d\tau}E(U)_a=0\,.
\eeq
Defining the angular velocity of the frame with respect to parallel transported axes
\beq
\frac{D E(U)_a}{d\tau} =\Omega(U) \times_U E(U)_a\,,
\eeq
namely
\begin{eqnarray}
\Omega(U)^1&=& E(U)_3\cdot \frac{D E(U)_2}{d\tau}\,,\nonumber\\
\Omega(U)^2&=&-E(U)_3\cdot \frac{D E(U)_1}{d\tau}\,,\nonumber\\
\Omega(U)^3&=& E(U)_2\cdot \frac{D E(U)_1}{d\tau}\,,
\end{eqnarray}
Eq.~\eqref{DUS} becomes
\beq
\label{DUS2}
\frac{dS^a}{d\tau}E(U)_a+\Omega(U) \times_U S=0\,.
\eeq
Reversing the sign of the angular velocity $\Omega(U)$ thus gives the angular velocity of the gyroscope spin vector with respect to the frame $\{E(U)_a\}$.
For example, if $E(U)_a$ is parallely transported along $U$, i.e., ${DE(U)_a}/{d\tau}=0$, then the above relation implies that the spin components $S^a$ are constant along $U$.

Now let $E(U)_a=E(U,u)_a$, for $u=m,n$, and let $S=S(U,u)^a E(U,u)_a$, so that
\beq
\label{DUS3}
\frac{dS(U,u)^a}{d\tau}+ \epsilon^{abc}\Omega(U,u)_b S(U,u)_c=0\,,
\eeq
with 
\beq\label{delUE}
\frac{D}{d\tau} E(U,u)_a =\Omega(U,u)\times_U E(U,u)_a\,.
\eeq
One may evaluate the angular velocity $\Omega(U,u)=\Omega(U,u)^aE(U,u)_a$ in terms of the relative motion of the geodesic $U$ and the observer family $u$ as a sum of the following three terms as given in  Ref.~\cite{Jantzen:1992rg}
\beq
\label{OmegaUu}
\Omega(U,u)=-\gamma(U,u)[\omega_{({\rm fw},u)}+\omega_{({\rm sc},U,u)}+\omega_{({\rm geo},U,u)}]\,,
\eeq
adopting its notation for the Fermi-Walker and the spatial curvature angular rotation vectors which characterize the covariant derivatives of the orthonormal frame along the orbit
\beq
P(u)\nabla_U e(u)_a = -\gamma(U,u)[\omega_{({\rm fw},u)}
+\omega_{({\rm sc},U,u)}]\times_{u}e(u)_a\,,\nonumber
\eeq
as well as the geodetic precession term in the gyroscope precession formula (see Eq.~(9.10) of 
Ref.~\cite{Jantzen:1992rg})
\beq
\label{omgUu}
\omega_{({\rm geo},U,u)}=\frac{1}{1+\gamma(U,u) } \,\nu(U,u)\times_u F^{(G)}_{({\rm fw},U,u)}\,,
\eeq
defined in terms of the spatial gravitational force 
$ F^{(G)}_{({\rm fw},U,u)} = -\nabla_U \,u$.
Here we are sloppily identifying the symbols for the various angular velocities which have the same orthonormal components before and after the boost from the observer frame to the geodesic frame.

Indeed this result for the angular velocity terms
Eq.~\eqref{OmegaUu} can be derived by first differentiating  along $U$ the relative boost $B_{\rm (lrs)}(U,u)$ relating $E(U,u)_a$ to $e(u)_a$ in \eqref{Blrs} to find
\begin{widetext}

\begin{eqnarray}
\label{come_si_trova_Omega}
\nabla_U E(U,u)_a &=& \nabla_U e(u)_a -\frac{\gamma(U,u)\nu(U,u)_a}{1+\gamma(U,u)}\,F^{(G)}_{({\rm fw},U,u)}
-\frac{\gamma^2(U,u)\nu(U,u)_a}{(1+\gamma(U,u))^2}\, (U+u)\,\nu(U,u)\cdot  F^{(G)}_{({\rm fw},U,u)}\nonumber\\
&&
+\frac{1}{1+\gamma(U,u)}\,(U+u)\,U\cdot \nabla_U e(u)_a\nonumber\\
&=& \mathcal{B} \rightcontract\left[\nabla_U e(u)_a 
-\frac{\gamma(U,u)\nu(U,u)_a}{1+\gamma(U,u)}\,F^{(G)}_{({\rm fw},U,u)}  \right]\,,
\end{eqnarray}
where $\mathcal{B}$ is the following tensor which extends the action of $B_{\rm (lrs)}(U,u)$ to vectors outside $LRS_u$
\beq
\mathcal{B}=I+\frac{(U+u)\otimes U^\flat}{1+\gamma(U,u)}\,,
\eeq
by mapping $U+u$ to 0, namely $\mathcal{B}\rightcontract (U+u)=0$.
One can then replace $\nabla_Ue(u)_a$ in the second line of Eq.~\eqref{come_si_trova_Omega} using 
\begin{eqnarray}
\nabla_U e(u)_a&=& P(u)\nabla_Ue(u)_a - [u\cdot \nabla_U e(u)_a] \, u
\nonumber\\
&=& -\gamma(U,u)[\omega_{({\rm fw},u)}+\omega_{({\rm sc},U,u)}]\times_{u}e(u)_a
-u F^{(G)}_{({\rm fw},U,u)}{}_a \,,
\end{eqnarray}
so that
\begin{eqnarray}
\label{come_si_trova_Omega2}
\frac{D}{d\tau} E(U,u)_a 
&=& \mathcal{B}\rightcontract \left[-\gamma(U,u)[\omega_{({\rm fw},u)}+\omega_{({\rm sc},U,u)}]\times_{u}e(u)_a
-u F^{(G)}_{({\rm fw},U,u)}{}_a 
-\frac{\gamma(U,u)\nu(U,u)_a}{1+\gamma(U,u)}F^{(G)}_{({\rm fw},U,u)}  \right]\nonumber\\
&=& \mathcal{B}\rightcontract \left[-\gamma(U,u)[\omega_{({\rm fw},u)}
+\omega_{({\rm sc},U,u)}+\omega_{({\rm geo},U,u)}]\times_{u}e(u)_a
-u F^{(G)}_{({\rm fw},U,u)}{}_a\right. \nonumber\\
&& \qquad \left.  
-\frac{\gamma(U,u)\nu(U,u)_a}{1+\gamma(U,u)}F^{(G)}_{({\rm fw},U,u)}+\gamma(U,u)\, \omega_{({\rm geo},U,u)} \times_{u}e(u)_a 
\vphantom{\frac{\gamma(U,u)\nu(U,u)_a}{1+\gamma(U,u)}} 
\right]\nonumber\\
&=& \mathcal{B}\rightcontract \left[\Omega(U,u)\times_{u}e(u)_a
-u F^{(G)}_{({\rm fw},U,u)}{}_a -\frac{\gamma(U,u)\nu(U,u)_a}{1+\gamma(U,u)}F^{(G)}_{({\rm fw},U,u)} +\gamma(U,u)\, \omega_{({\rm geo},U,u)} \times_{u}e(u)_a \right]\,,
\end{eqnarray}
where  in the second line we have added and subtracted the term (using the triple cross product identity on \eqref{omgUu})
\begin{eqnarray}
\label{last_piece}
\gamma(U,u)\, \omega_{({\rm geo},U,u)} \times_{u}e(u)_a
= -\frac{\gamma(U,u)}{1+\gamma(U,u)} \left[\nu(U,u) F^{(G)}_{({\rm fw},U,u)}{}_a   
      - F^{(G)}_{({\rm fw},U,u)}\nu(U,u)_a\right]
\end{eqnarray} 
to form the combination $\Omega(U,u)\times_{u}e(u)_a$.
Expanding the last term in Eq. \eqref{come_si_trova_Omega2} by using Eq. \eqref{last_piece} leads to simplifications (the term proportional to the vector $F^{(G)}_{({\rm fw},U,u)}$ disappears), so that
combining the various pieces  we find
\begin{eqnarray}
\label{come_si_trova_Omega3}
\frac{D}{d\tau} E(U,u)_a 
&=& \mathcal{B}\rightcontract \left[\Omega(U,u)\times_{u}e(u)_a-\frac{U+u}{1+\gamma(U,u)} F^{(G)}_{({\rm fw},U,u)}{}_a \right]\nonumber\\
&=& \mathcal{B}\rightcontract \left[  \Omega(U,u)\times_{u}e(u)_a  \right]\nonumber\\
&=& [\mathcal{B}\Omega(U,u)]\times_{U}E(U,u)_a\,,
\end{eqnarray}
since the final term proportional to $U+u$ is mapped onto $0$ by $\mathcal{B}$. Note that in the second line of Eq. \eqref{come_si_trova_Omega3} we have 
$\Omega(U,u)= \Omega(U,u)^ae(u)_a$, whereas in the last line $\mathcal{B}\Omega(U,u)=\Omega(U,u)^a\, E(U,u)_a$ denotes the 
boosted vector, which has the same components, but referred to the boosted frame $E(U,u)_a$.

\end{widetext}

The following sections will consider explicitly the orthonormal triads in $LRS_U$ associated with the static observers ($u=m$), ZAMOs ($u=n$) and the Frenet-Serret frame along $U$ involved in  Marck's construction. These triads are all related to each other by relative rotations whose time derivatives along the geodesic produce their relative angular velocities.
Finally, a parallely propagated triad along $U$ can also be identified by Marck's additional planar rotation of the Frenet-Serret triad.
Studying each of these frames along a generic geodesic requires rather involved computations which often hide their geometrical content.

\section{Gyroscope precession and static observers}

Specializing the angular velocity \eqref{OmegaUu}  to the case $u=m$, 
 a lengthy calculation leads to the explicit result

\begin{widetext}

\begin{eqnarray}
\label{Omega}
\Omega(U,m)^1&=&
\frac{\gamma^2}{(1+\gamma)E^2\Sigma^{3/2}\Delta^{1/2}}\left\{
aM\sin\theta(r^2-a^2\cos^2\theta)\dot r \dot \theta
+\frac{\cos\theta}{\Sigma^2\sin^2\theta}\bigg[
Ex[\Sigma^3-2Mr(\Sigma^2-2a^4\cos^2\theta\sin^2\theta)]\right.\nonumber\\
&&
+aE^2[\Sigma^3-2Mr(\Sigma^2-a^4\sin^2\theta\cos^4\theta)]
+2Mra\sin^2\theta(a^2x^2-\Sigma\Delta\sinh^2\beta)\nonumber\\
&&\left.
+\frac{E\Sigma}{\gamma}\left[
x(\Sigma^2-2Mr(\Sigma-a^2\sin^2\theta))
+aE(\Sigma^2-2Mr(\Sigma-a^2\sin^2\theta\cos^2\theta))
\right]
\bigg]
\right\}
\,,\nonumber\\
\Omega(U,m)^2&=&
\frac{\gamma^2}{(1+\gamma)E^2\Sigma^{3/2}}\left\{
2aMr\cos\theta\,\dot r \dot \theta
+\frac{1}{\Sigma^2\Delta\sin\theta}\bigg[
Ex[(r-M)\Sigma^2(\Sigma-2Mr)-2Mr^2\Delta(r^2-a^2\cos^2\theta)]\right.\nonumber\\
&&
+arE^2[\Sigma^3-Mr(\Sigma^2(1+\cos^2\theta)+\cos^2\theta\Delta(\Sigma+2r^2))]
+aM\Delta(r^2-a^2\cos^2\theta)(x^2-\Sigma\sin^2\theta\cosh^2\beta)\nonumber\\
&&\left.
+\frac{E\Sigma}{\gamma}\left[
x((r-M)\Sigma^2-M\Sigma(r^2-a^2)-2Mr^2\Delta)
+arE(\Sigma^2-Mr(\Sigma(1+\cos^2\theta)+2\cos^2\theta\Delta))
\right]
\bigg]
\right\}
\,,\nonumber\\
\Omega(U,m)^3&=&
\frac{a\cos\theta\,\dot r}{E\,\Sigma^2\Delta^{1/2}\sin\theta}\left[
aE\Sigma\sin^2\theta
+\frac{2Mr\gamma}{1+\gamma}(x+aE\cos^2\theta)
\right]\nonumber\\
&&
+\frac{\dot \theta}{\Sigma}\left[
r\Delta^{1/2}
+\frac{M\gamma}{(1+\gamma)E\Sigma\Delta^{1/2}}(r^2-a^2\cos^2\theta)(ax-r^2E)
\right]
\,,
\end{eqnarray}
with $\gamma = E/\sqrt{1-2Mr/\Sigma}$.
On the equatorial plane we have $\Omega(U,m)^1=0=\Omega(U,m)^3$ and 
\beq
\label{Omegab2equat}
\Omega(U,m)^2\vert_{\theta=\pi/2}=
-\frac{\gamma}{E^2r^3\Delta}\left[
\frac{M\Delta}{1+\gamma}(Ex+a)+Er^3(x+aE)-aM(\Delta+E^2r^2)-MEx(2r^2-a^2+3\Delta)
\right]
\,,
\eeq

\end{widetext}
with $\gamma = E/\sqrt{1-2M/r}$.
It is important to recall that this represents the angular velocity of the axes $E(U,m)_a$ with respect to parallel transported axes along the world line, while the angular velocity of the parallel transported spin vector of a gyroscope with respect to these axes has the opposite sign.
The previous expressions simplify considerably for special orbits, e.g., spherical orbits ($\dot r=0$) or orbits with constant polar angle ($\dot\theta=0$).
These cases will be studied below.

\section{Gyroscope precession and ZAMOs}

One may also express the angular velocity  \eqref{DUS3} of a parallel transported gyroscopic spin vector $S=S(U,n)^aE(U,n)_a$ relative to the boosted ZAMO frame in $LRS_U$ with a similar decomposition into geodetic, Fermi-Walker and space curvature terms. These are given explicitly 
 in Appendix A.

The boosted ZAMO frame vectors $\{E(U,n)_a\}$ are then related to the static observer frame vectors $\{E(U,m)_a\}$ by a spatial rotation in LRS$_U$, i.e.,
\beq
\label{static_to_zamo}
E(U,m)_a=E(U,n)_b {\sf R}^b{}_a\,.
\eeq
The spatial rotation ${\sf R}$  is just the Wigner rotation due to the combination of initial azimuthal boost from the ZAMOs to the static observers 
\beq
e(m)_a=B_{\rm (lrs)}(m,n)\,  e_{\hat a}
\eeq
with the successive boost from $LRS_m$ to $LRS_U$. 
Applying $B_{\rm (lrs)}(U,m)$ to both sides then leads to
\begin{eqnarray}
E(U,m)_a&=&B_{\rm (lrs)}(U,m)\,e(m)_a\nonumber\\
&=&B_{\rm (lrs)}(U,m)\,B_{\rm (lrs)}(m,n)\,  e_{\hat a}\nonumber\\
&=&{\sf R}^b{}_a\,B_{\rm (lrs)}(U,n)\,e_{\hat b}\nonumber\\
&=&E(U,n)_b{\sf R}^b{}_a\,.
\end{eqnarray}

We conclude this section by noting that the above ZAMO boosted frame is just the one underlying the construction of the Hamiltonian of a spinning particle in a curved spacetime to linear order in spin due to Barausse, Racine and Buonanno~\cite{Barausse:2009aa}. 
Unfortunately, the Hamiltonian formalism loses contact with the geometrical content of the problem. Taking advantage of the relative observer point of view as well as the associated spacetime splitting techniques, it is easy to show that their explicit expressions (5.35)--(5.37) for the spin precession angular velocity components reduce to 
\begin{eqnarray}
\label{calHabrb}
{\bar H}^{\hat a}&=&N\left[\omega_{({\rm geo},U,n)}+\omega_{({\rm fw},n)}+\omega_{({\rm sc},U,n)}\right]^{\hat a}\nonumber\\
&=&-\frac{N}{\gamma(U,n)}\Omega(U,n)^a
\,,
\end{eqnarray}
with the replacements $\sqrt{Q}=\gamma(U,n)$ and ${\hat P}_a=\gamma(U,n)\nu(U,n)^{\hat a}\sqrt{g_{aa}}$ in their corresponding equations to first order in spin.

\section{Gyroscope precession and the degenerate Frenet-Serret frame}

Another possible choice for a natural frame along $U$ involves the Frenet-Serret procedure, which is a degenerate case, since $U$ is geodesic.
The construction of such a frame is intimately linked to Marck's construction of a parallely propagated frame, which in the case of equatorial plane motion gives a direct route to evaluating the gyro spin precession. In the general case the geometric construction is more involved and
requires a rather involved treatment consisting of the following main steps.

\subsection{Step 1: decomposing $U$ in Carter's frame}

Consider a generic timelike geodesic with unit tangent vector \eqref{Ugeogen} decomposed relative to the Carter observers as in Eq.~\eqref{Uucar}.
To make the notation less cumbersome below we introduce the abbreviations $\gamma(U,u_{\rm (car)})=\gamma_{\rm c}$, $\nu(U,u_{\rm (car)})=\nu_{\rm c}$.
For later use let us introduce the angular part $\nu^\top$ of the Carter relative velocity and an orthogonal vector $\nu^\perp=  E(u_{\rm (car)}) \times_{u_{\rm (car)}} \nu^\top $ of the same magnitude in the angular subspace
\begin{eqnarray}
\nu^\top &=& \nu_{\rm c}^2  E(u_{\rm (car)})_2+ \nu_{\rm c}^3  E(u_{\rm (car)})_3\equiv ||\nu^\top ||  \hat \nu^\top\,, \nonumber\\
\nu^\perp &=& -\nu_{\rm c}^3  E(u_{\rm (car)})_2+\nu_{\rm c}^2  E(u_{\rm (car)})_3\equiv ||\nu^\perp ||  \hat \nu^\perp\,, \nonumber\\
\end{eqnarray}
with $||\nu^\top ||=||\nu^\perp || =\sqrt{[\nu_{\rm c}^2]^2+[\nu_{\rm c}^3]^2}$.
The orthogonal decomposition into radial and angular directions in the Carter frame is the starting point for solving the equations of parallel transport along a geodesic, as done by Marck and explained geometrically for the simpler case of equatorial plane geodesics in our previous articles
\cite{Bini:2016iym,Bini:2016ovy}.

\subsection{Step 2: boosting Carter's frame along the radial direction}

Using the components of $U$ with respect to the Carter observers we define a frame $\{E^{\rm rad}_\alpha\}$ in which we first boost along the radial direction by the radial component of the relative 4-velocity of the gyro to obtain a new radially comoving radial direction in a new local rest space, and then pick the next frame vector to be along the direction of the remaining angular component of the Carter relative velocity,  and then the final angular axis orthogonal to the first one forming a right-handed spatial frame within $LRS_{E^{\rm rad}_0}$
\begin{eqnarray}
E^{\rm rad}_0 &=&\gamma^\Vert [u_{\rm (car)}+\nu_{\rm c}^1 E(u_{\rm (car)})_1 ]\,,\nonumber\\
E^{\rm rad}_1 &=&\gamma^\Vert [\nu_{\rm c}^1 u_{\rm (car)}+ E(u_{\rm (car)})_1]
\equiv B(\alpha) E(u_{\rm (car)})_1 \,,\nonumber\\
E^{\rm rad}_2 &=& \hat \nu^\top  \,,\qquad
E^{\rm rad}_3 =\hat \nu^\perp\,,
\end{eqnarray}
where the boost rapidity $\alpha$ and gamma factor are defined by
\beq
\nu_{\rm c}^1=\tanh\alpha\,,\qquad
\gamma^\Vert=\cosh\alpha
=\frac{P}{\sqrt{\Delta(r^2+K)}}\,.
\eeq
Introducing spherical coordinates to parametrize the direction of the relative velocity in $LRS_{u_{\rm(car)}}$ with pole along the radial velocity direction
\beq
  \langle \hat\nu_c^1,\hat\nu_c^2,\hat\nu_c^3\rangle 
   = \langle \cos\Theta,\sin\Theta \cos\Phi,\sin\Theta\sin\Phi \rangle\,,
\eeq
then the rotation to the intermediate frame adapted to the Carter relative velocity decomposition is 
\begin{widetext}
\beq
\begin{pmatrix}
E_1^{\rm rad}& E_2^{\rm rad} & E_3^{\rm rad}
\end{pmatrix}
= \begin{pmatrix}
B(\alpha) E(u_{\rm(car)})_1 & E(u_{\rm(car)})_2 & E(u_{\rm(car)})_3
\end{pmatrix}
\begin{pmatrix}
1 & 0 & 0 \\
0 & \cos\Phi & -\sin\Phi \\
0 & \sin\Phi & \cos\Phi
\end{pmatrix}
\,.
\eeq
\end{widetext}

The geodesic 4-velocity in this frame then has the explicit form (boosting along the Carter angular relative velocity)
\beq\label{Ubetarad}
U=\cosh \beta\, E^{\rm rad}_0+\sinh \beta\, E^{\rm rad}_2\,,
\eeq
where
\beq
\cosh \beta =\sqrt{\frac{K+r^2 }{\Sigma}}\,,\qquad \sinh \beta =  \sqrt{\frac{K-a^2\cos^2\theta }{\Sigma}}\,.
\eeq
From this relation one easily identify the orthogonal (spatial) direction in this plane 
\beq\label{e1def}
e_3=\sinh \beta\, E^{\rm rad}_0+\cosh \beta\, E^{\rm rad}_2\,.
\eeq

\subsection{Step 3: the successive boost in  the angular direction to a degenerate Frenet-Serret frame along $U$}

The final frame adapted to $U$ (i.e., to the gyro world line) is then obtained by boosting this intermediate frame $\{E_a^{\rm rad}\}$ into $LRS_U$ by a boost
along the direction of the angular relative motion, namely extending the relation \eqref{Ubetarad} to the full corresponding relative observer boost, which leaves the remaining two frame vectors $\{E_1^{\rm rad},E_3^{\rm rad}\}$ invariant
\beq
\begin{pmatrix}
U & e_3
\end{pmatrix}
=\begin{pmatrix} 
E_0^{\rm rad}& E_2^{\rm rad}\end{pmatrix}
\begin{pmatrix}
\cosh \beta & \sinh \beta \cr
\sinh \beta & \cosh \beta
\end{pmatrix}\,.
\eeq
Thus $e_3$ is aligned with the direction of the angular relative motion between the Carter and gyro local rest spaces,  reducing to the azimuthal direction in the equatorial plane case.

Marck showed that a unit vector $e_2$ orthogonal to both $U$ and $e_3$ which is also parallel propagated along $U$ arises naturally by normalizing the contraction of $U$ with the Killing-Yano 2-form of the Kerr spacetime. Because this 2-form is so simply expressed in both the Carter and intermediate frames (see Appendix A of Ref.~\cite{Bini:2016iym}), the resulting vector frame components
are obtained by a simple anisotropic rescaling of the two vector components of $U$ expressed in the form \eqref{Ubetarad}
\begin{eqnarray}
e_2&=&\frac{a\cos \theta}{\sqrt{K}} \cosh \beta\, E^{\rm rad}_1
-\frac{r}{\sqrt{K}} \sinh \beta\, E^{\rm rad}_3\nonumber\\
&\equiv&-(-\sin\Xi\, E^{\rm rad}_1+\cos\Xi\, E^{\rm rad}_3)
\,,
\end{eqnarray}
where
\beq
  \cos\Xi= \frac{r}{\sqrt{K}}\,\sinh\beta\,,\quad
  \sin\Xi= \frac{a\cos \theta}{\sqrt{K}}\,\cosh\beta\,.
\eeq
The last frame vector $e_1=e_2\times_U e_3$
 is then determined by orthogonality to be
\begin{eqnarray}
e_1&=&\frac{r}{\sqrt{K}} \sinh \beta\, E^{\rm rad}_1
   +\frac{a\cos \theta}{\sqrt{K}} \cosh \beta\, E^{\rm rad}_3\nonumber\\
&\equiv& \cos\Xi\,E^{\rm rad}_1+\sin\Xi\,E^{\rm rad}_3\,.
\end{eqnarray}
Together the two new frame vectors are obtained by a rotation in the plane orthogonal to the Carter angular relative velocity (modulo the time direction and a spatial reflection) to align the frame with the parallel transported vector $e_2$.

Note that in the equatorial plane where $\theta=\pi/2$ and $K=x^2$, then
 $\Xi=0$ and 
\beq
\cosh \beta =\frac{\sqrt{x^2+r^2}}{r}\,,\qquad 
\sinh \beta =  \frac{|x|}{r}\,,
\eeq
implying that $e_2={\rm sgn}(x) e_{\hat\theta}$ and
\begin{eqnarray}
e_1&=&\frac{r}{\sqrt{r^2+x^2}}\left[\frac{r^2+a^2}{\Delta}\dot r \,\partial_t
+\frac{P}{r^2}\,\partial_r
+\frac{a}{\Delta}\dot r\,\partial_\phi\right]
\,,\nonumber\\
e_3&=&\frac{|x|}{\sqrt{r^2+x^2}}\left\{\frac1{r^2}\left[\frac{a}{x}(r^2+x^2)+\frac{P}{\Delta}(r^2+a^2)\right]\partial_t\right.\nonumber\\
&&\left.
+\dot r \,\partial_r
+\frac1{r^2}\left[\frac{1}{x}(r^2+x^2)+a\frac{P}{\Delta}\right]\partial_\phi
\right\}
\,.
\end{eqnarray}
For simplicity in what follows, we will assume ${\rm sgn}(x)=1$ when taking the equatorial plane limit of general expressions, although one can easily take into account the case of negative $x$.

This frame $\{U,e_1,e_2,e_3\}$ is a degenerate Frenet-Serret frame along $U$, such that
\begin{eqnarray}\label{FSframeDEs}
\frac{DU}{d\tau}&=& 0\,,\qquad
\frac{De_1}{d\tau} = {\mathcal T} e_3\,,\nonumber\\
\frac{De_2}{d\tau} &=& 0 \,,\qquad 
\frac{De_3}{d\tau} = -{\mathcal T} e_1\,,
\end{eqnarray}
with 
\begin{eqnarray}
\label{cal_T_torsion}
{\mathcal T}&=&\frac{\sqrt{K}}{\Sigma}\left[ \frac{P}{r^2+K}+\frac{aB}{K-a^2\cos^2\theta} \right]\nonumber\\
&=&\frac{\sqrt{K}}{\Sigma^2}\left[ \frac{P}{\cosh^2\beta}+\frac{aB}{\sinh^2\beta} \right]\\
&=& -\sqrt{K\Sigma} \left[\frac{\sqrt{\Delta}}{r^2+K} u_{\rm (car)}-\frac{a\sin \theta}{K-a^2\cos^2\theta}\bar u_{\rm (car)}  \right]\cdot U\,,\nonumber
\end{eqnarray}
the only surviving (spacetime) torsion of the world line.

Note that by normalizing the timelike vector in square brackets in the final line of Eq.~\eqref{cal_T_torsion} we can identify a new timelike vector $u'$ 
in the time-azimuthal plane of the tangent space
such that
$\mathcal{T}=-\T \, u'\cdot U =\T   \, \gamma(U,u')$, 
namely
\beq
u'=\gamma(u',u_{\rm (car)})\left[ u_{\rm (car)}+\nu(u',u_{\rm (car)}) \bar u_{\rm (car)}  \right]\,,
\eeq
where
\begin{eqnarray}
\nu(u',u_{\rm (car)})&=&  
-\frac{K+r^2}{K-a^2\cos^2\theta}\frac{a\sin \theta}{\sqrt{\Delta}}
\,,
\end{eqnarray}
and
\beq
\T =\frac{\sqrt{K\Sigma\Delta}}{(r^2+K)\gamma(u',u_{\rm (car)})}\,.
\eeq

The corresponding angular velocity vector is 
$\Omega_{\rm(FS)} = - \mathcal{T} e_2$, so that
\beq\label{omegaT}
  \frac{De_a}{d\tau} = \Omega_{\rm(FS)}\times_U e_a\,.
\eeq
We then find for $S=S_{\rm(FS)}^ae_a$
\beq
\frac{dS_{\rm(FS)}^a}{d\tau}e_a+ \Omega_{\rm(FS)}\times_U S=0\,.
\eeq

\section{Gyroscope precession and parallel propagated frame}

The vectors $e_1$ and $e_3$ were found by Marck ``by inspection" of generic conditions determining vectors in the plane orthogonal to $U$ and $e_2$, and in contrast with $e_2$ are not parallel propagated along $U$. To get a frame $\{E_i\}$ (with $E_2=e_2$) that is parallel propagated, it is enough to rotate them by an appropriate angle $\Psi$, 
\beq\label{K30}
E_1= e_1 \cos \Psi - e_3 \sin \Psi\,,\ 
E_3= e_1 \sin \Psi +e_3 \cos \Psi\,,
\eeq
such that ~\cite{marck1,marck2}
\beq
\label{dpsi_dtau_0}
\frac{d\Psi}{d\tau}={\mathcal T}\,.
\eeq
Eq.~\eqref{dpsi_dtau_0} must be then integrated along the orbit to determine $\Psi$ modulo an arbitrary initial value. 

The components of the spin vector with respect to that frame $S=S_{\rm(par)}^aE_a$ are constant, namely
\beq
\frac{dS_{\rm(par)}^a}{d\tau}=0\,.
\eeq

\section{Gyroscope precession with respect to Cartesian axes at spatial infinity}

In the previous sections the precession of a gyroscope has been computed with respect a triad of axes defined all along its world line through comparison with the boost of a spherical coordinate triad associated with either the static observers or the ZAMOs, or by using the Frenet-Serret frame $\{e_a\}$, all of which are adapted to the local rest space of the gyro along its orbit and related to each other by time-dependent rotations whose time derivatives determine their relative angular velocities.
In the first two cases as the gyro moves in the spherical coordinate grid in a nonradial direction,
the directions of these triads with respect to spatial infinity rotate. The same is true for the Frenet-Serret frame since its construction relies on the spherical decomposition of the relative velocity.

To measure rotation with respect to local directions which are fixed with respect to spatial infinity, we need local static Cartesian-like axes whose directions with respect to spatial infinity along the world line do not change. Motion relative to the static observers thus leads to an orbital angular velocity contribution to the precession of the gyro spin with respect to these latter axes due to its motion relative to those observers.
The boost from $LRS_m$ to $LRS_U$ removes the aberration of those directions.
An additional Wigner rotation contributes for comparison with boosted ZAMO axes due to the additional relative boost, while the Frenet-Serret frame has an additional rotation explicitly given below. Thus a crucial rotation from Cartesian to spherical directions must be taken into account, leading to an ``orbital" contribution to the angular velocity of a parallel transported spin vector.

Consider the family of  static observers which exist only in the spacetime region outside the black hole ergosphere where $g_{tt}<0$. The static observer spherical frame of \eqref{static_triad} is locked onto the spatial coordinate grid dragged along by the static observers, but one can instead lock a triad onto the distant Cartesian coordinates associated with the Boyer-Lindquist coordinates by locally rotating the spherical axes to align them with axes pointing to fixed directions at spatial infinity.

This Cartesian-like orthonormal frame $\{e_A\},A=x,y,z$  in the local rest space $LRS_{m}$ is defined by the same rotation as in flat space spherical coordinates
\begin{eqnarray}
\begin{pmatrix} e(m)_1 & e(m)_2 & e(m)_3 \end{pmatrix} 
=\begin{pmatrix} e_x & e_y & e_z \end{pmatrix} \,
\mathcal{R}(\theta,\phi)
\end{eqnarray}
or
\beq\label{exyz}
e(m)_a =e_A \mathcal{R}(\theta,\phi)^A{}_a\,,
\eeq
where 
\beq\label{orbitalrotation}
\mathcal{R}(\theta,\phi)
=
\begin{pmatrix}
& \sin\theta\cos\phi &\cos\theta\cos\phi & -\sin\phi\\
& \sin\theta\sin\phi &\cos\theta\sin\phi &\cos\phi\\
& \cos\theta         &-\sin\theta        & 0
\end{pmatrix}
\eeq
is the usual flat space rotation matrix relating the Cartesian and spherical coordinate orthonormal frames. 
To show 3d plots of numerical geodesic paths in space, it is convenient to introduce the corresponding Cartesian coordinates
\beq\label{xyz}
x = r\sin\theta \cos\phi\,,\quad
y = r\sin\theta \sin\phi\,,\quad
z=r \cos\theta\,,
\eeq
and for polar geodesics described below it is convenient to allow all real values for $\theta$ along them through this representation.

The frame $\{e_A\}$ has its orientation fixed with respect to the ``distant stars." Then the following derivative defines proper time rate of change of the relative rotation
\beq
  {\mathcal{R}}^{-1}\,\dot{\mathcal{R}}
 = \begin{pmatrix}
& 0       &-\dot\theta      & -\sin\theta\,\dot\phi\\
& \dot\theta & 0            & -\cos\theta\,\dot\phi\\
& \sin\theta\,\dot\phi &\cos\theta\,\dot\phi & 0
\end{pmatrix} 
\equiv
 \Omega_{\rm{(orb)}}^{i} {L}_i
\,,
\eeq
where the three antisymmetric matrix generators of the active action of the rotation group are defined by $[L_i]^j{}_k = \epsilon _{jik}$ in terms of the Levi-Civita symbol and
\begin{eqnarray}
\label{omegaorb}
  {\Omega}_{\rm(orb)} 
&=&\Omega_{\rm(orb)}^{\hat r}e(m)_1+\Omega_{\rm(orb)}^{\hat \theta}e(m)_2+\Omega_{\rm(orb)}^{\hat \phi} e(m)_3\nonumber\\
&=&  \cos\theta \,\dot\phi \,e(m)_1
     -\sin\theta\,\dot\phi \,e(m)_2\nonumber\\
&&
     +\dot\theta  \,e(m)_3\,.
\end{eqnarray}
Then the proper time derivative of components with respect to the spherical frame of any vector defined along the gyroscope world line are related to those with respect to the Cartesian-like frame $\{e_A\}$
by the following ``orbital" angular velocity
\beq 
   \dot S^A = (\mathcal{R}^A{}_i S^i)\dot{\phantom{\i}}
 = \mathcal{R}^A{}_i(\dot S{}^i+\epsilon_{ijk} \Omega_{\rm(orb)}^j S^k)\,.
\eeq
This describes the orbital angular velocity along a geodesic of the static observer spherical frame relative to the distantly nonrotating celestial sphere, locally represented by the static observer Cartesian-like frame. 

Boosting both the spherical frame to $\{B(U,m) e(m)_a\}$
and the Cartesian-like frame to $ \{B(U,m) e_A\}$, the same rotation applies either before or after the boost, so the angular velocity has the same components in the boosted frame
\begin{eqnarray}
{\Omega}_{\rm(orb)}(U,m)
&=&\cos\theta \,\dot\phi\, E(U,m)_1 - \sin\theta\,\dot\phi\, E(U,m)_2\nonumber\\
&& + \dot\theta\, E(U,m)_3\,.
\end{eqnarray}
A gyroscope moving along its world line then  precesses with respect to these latter axes in $LRS_U$ by the angular velocity 
\beq
\label{Omegaprec}
{\Omega}_{\rm(prec)}={\Omega}_{\rm(orb)}(U,m)-{\Omega}(U,m)\,,
\eeq
since the angular velocity of the spin vector with respect to the boosted spherical axes is $-{\Omega}(U,m)$.

\begin{figure*}
\[
\begin{array}{cc}
\includegraphics[scale=0.35]{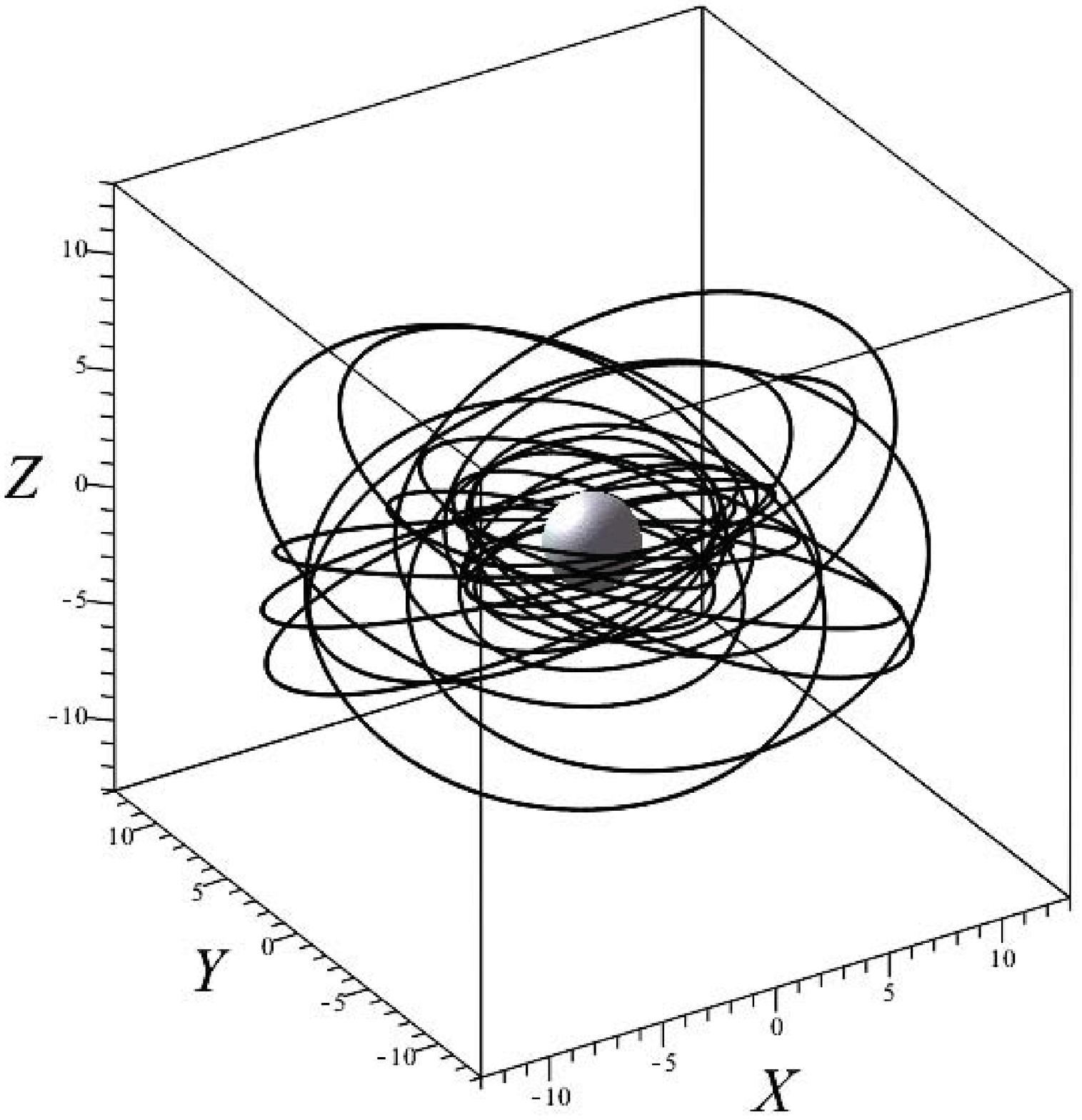}& 
\includegraphics[scale=0.35]{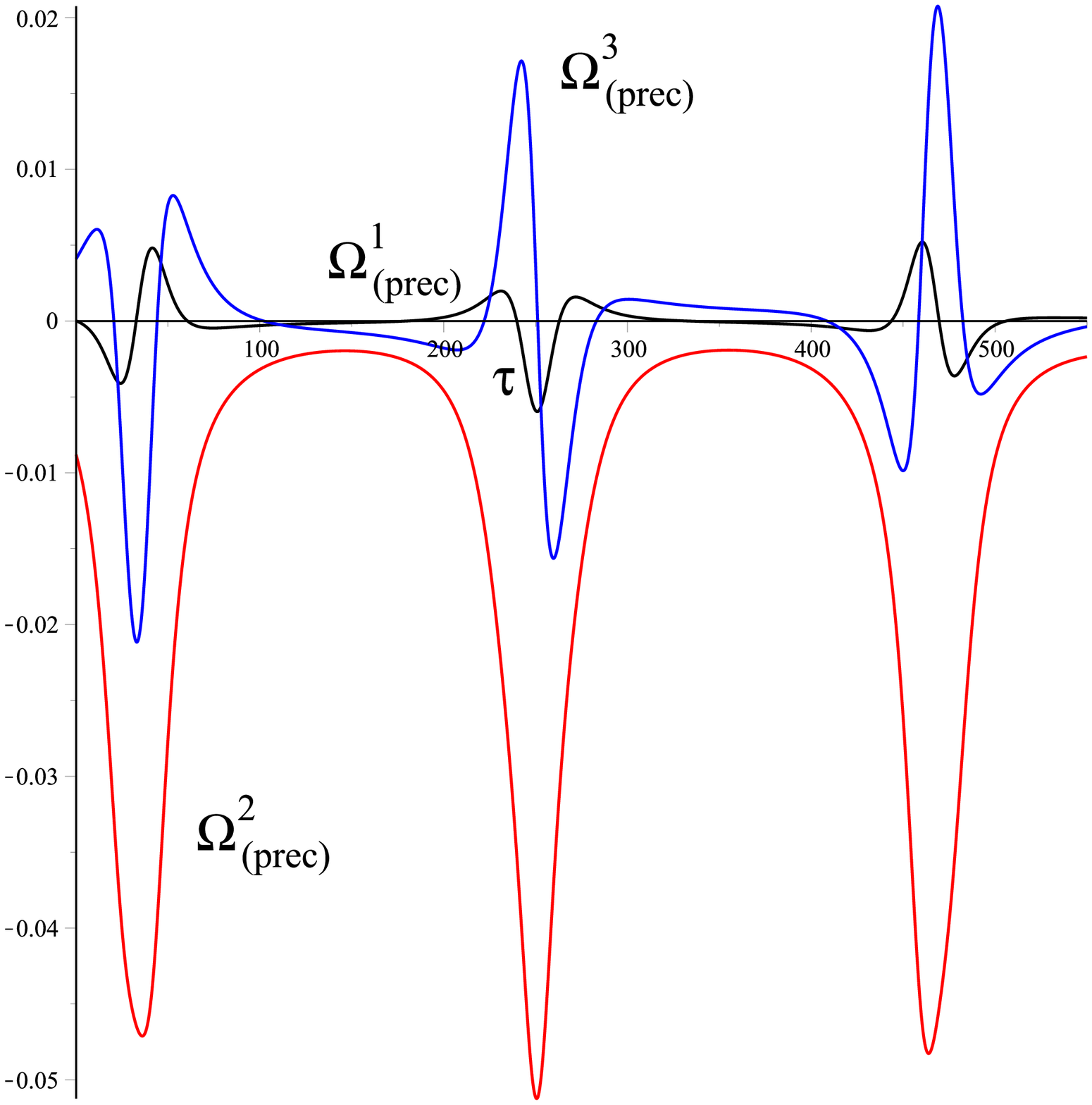}\cr
(a) & (b)
\end{array}
\]
\caption{An example of a general bound geodesic is shown in panel (a). 
The black hole parameters are chosen so that $M=1$ and $a/M=0.5$. The orbital parameters following initial values are given by $E=0.95$, $L/M=3$, $K/M^2=8$ with motion starting at $\tau=0$ from the point ($r_0/M=8$, $\theta_0=\pi/2$, $\phi_0=0$) radially ingoing ($\epsilon_r=-1$, $\dot r(0)\approx-0.1314$) and $\theta$ increasing ($\epsilon_\theta=1$, $\dot \theta(0)\approx0.01991$). 
The $r$-motion is confined between $r_{\rm min}/M\approx4.7451$ and $r_{\rm max}/M\approx13.1587$, whereas the $\theta$-motion between $\theta_{\rm min}\approx1.1695$ and $\theta_{\rm max}=\pi-\theta_{\rm min}\approx1.9721$. 
The corresponding evolution of the three components (with respect to the frame $\{E(U,m)_i\}$) of the gyroscope precession angular velocity ${\Omega}_{\rm(prec)}$ along the orbit are shown in panel (b) for about five revolutions.
}
\label{fig:1}
\end{figure*}

\begin{figure*}
\[
\begin{array}{cc}
\includegraphics[scale=0.35]{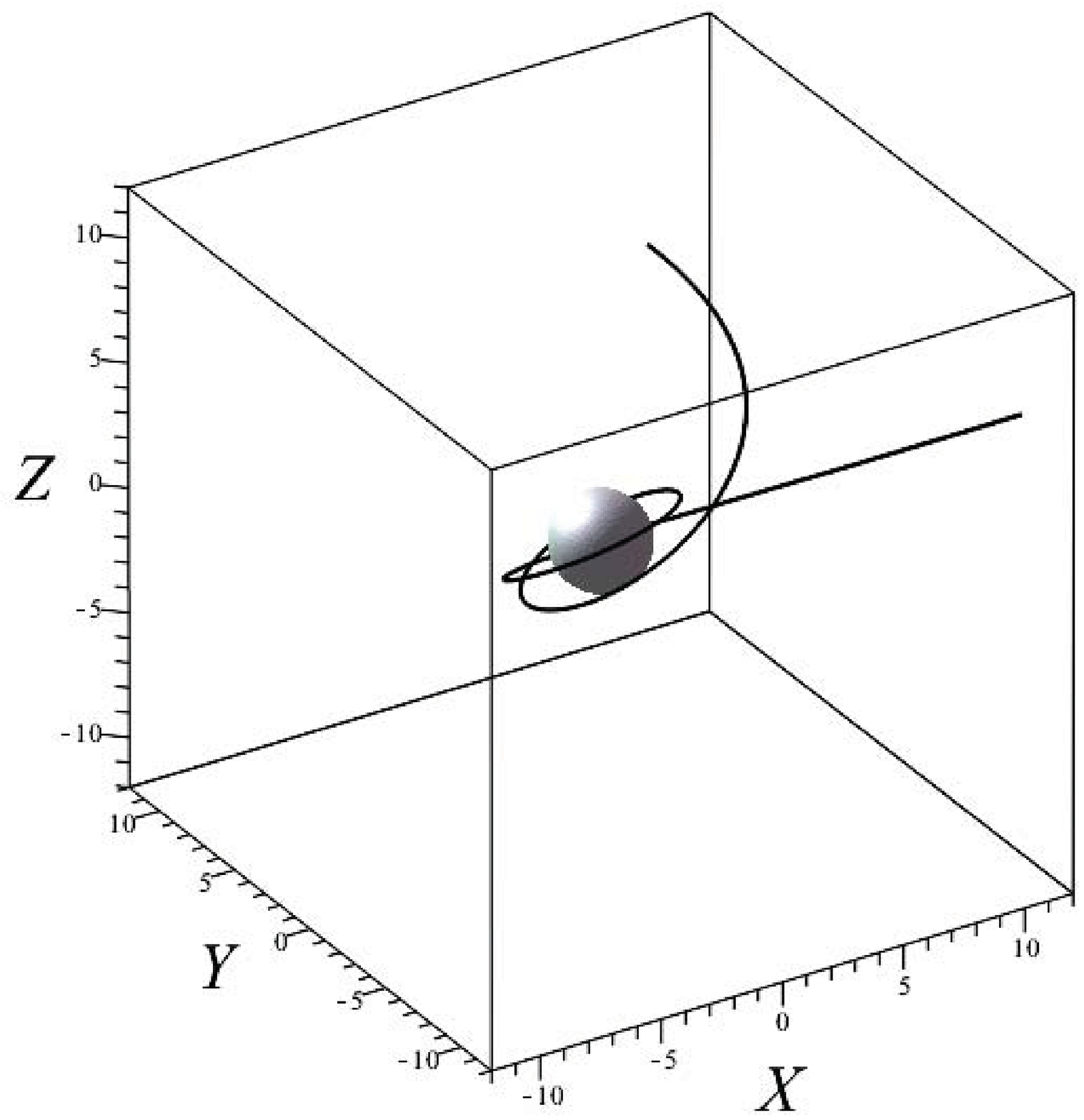}& 
\includegraphics[scale=0.35]{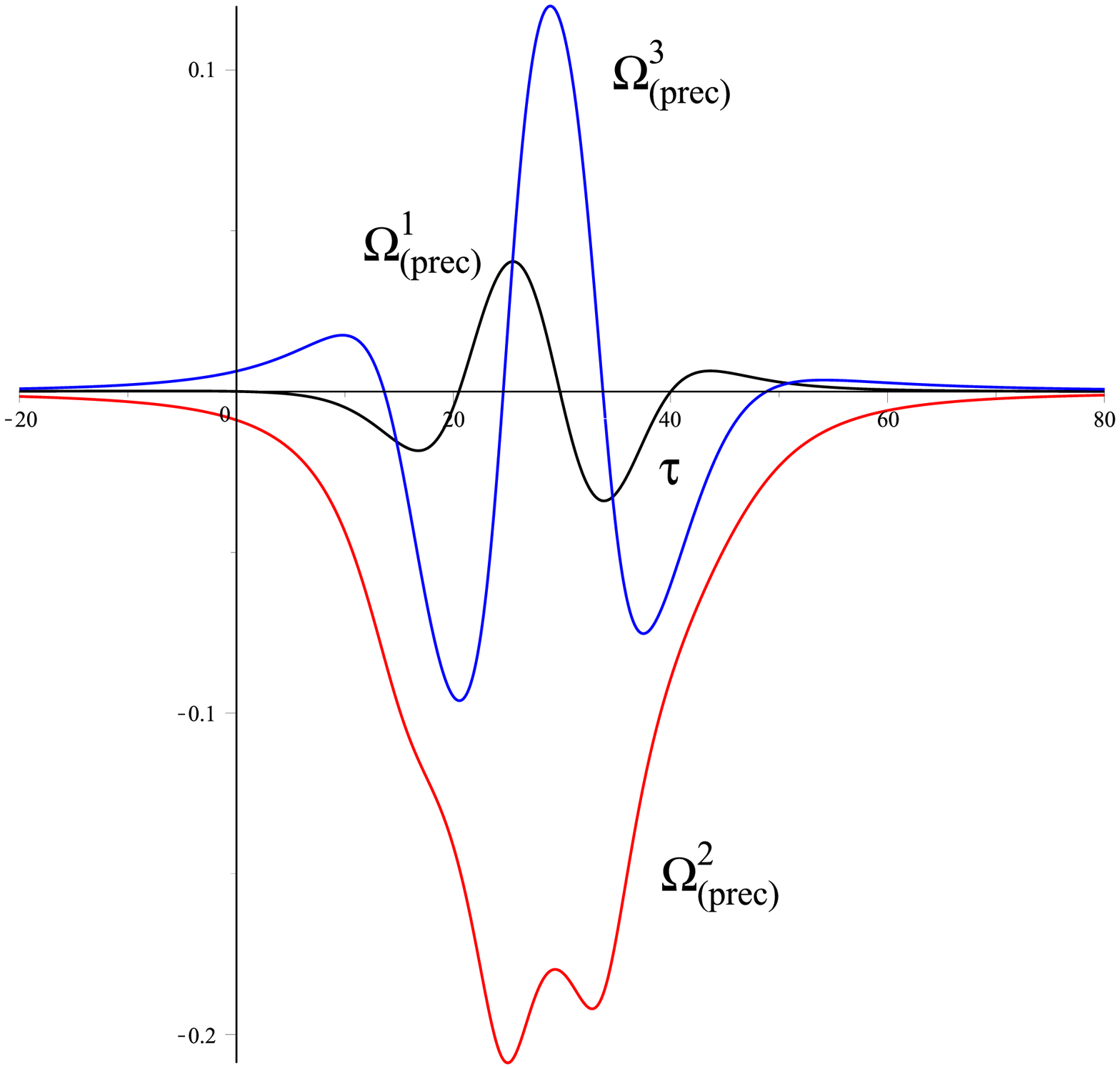}\cr
(a) & (b)
\end{array}
\]
\caption{An example of general unbound geodesic is shown in panel (a). 
The black hole parameters are chosen so that $M=1$ and $a/M=0.5$. The orbital parameters following initial values are given by $E=0.9$, $L/M=3$, $K/M^2=10.5$ with motion starting at $\tau=0$ from the point ($r_0/M=8$, $\theta_0=\pi/2$, $\phi_0=0$) radially ingoing ($\epsilon_r=-1$, $\dot r(0)\approx-0.3210$) and $\theta$ increasing ($\epsilon_\theta=1$, $\dot \theta(0)\approx0.03124$). 
The numerical integration of the geodesic equations is completed forward and backward in proper time in order to cover the whole scattering process. 
The corresponding evolution of the components of the gyroscope precession frequency along the orbit are shown in panel (b).
}
\label{fig:2}
\end{figure*}

\begin{figure*}
\[
\begin{array}{cc}
\includegraphics[scale=0.35]{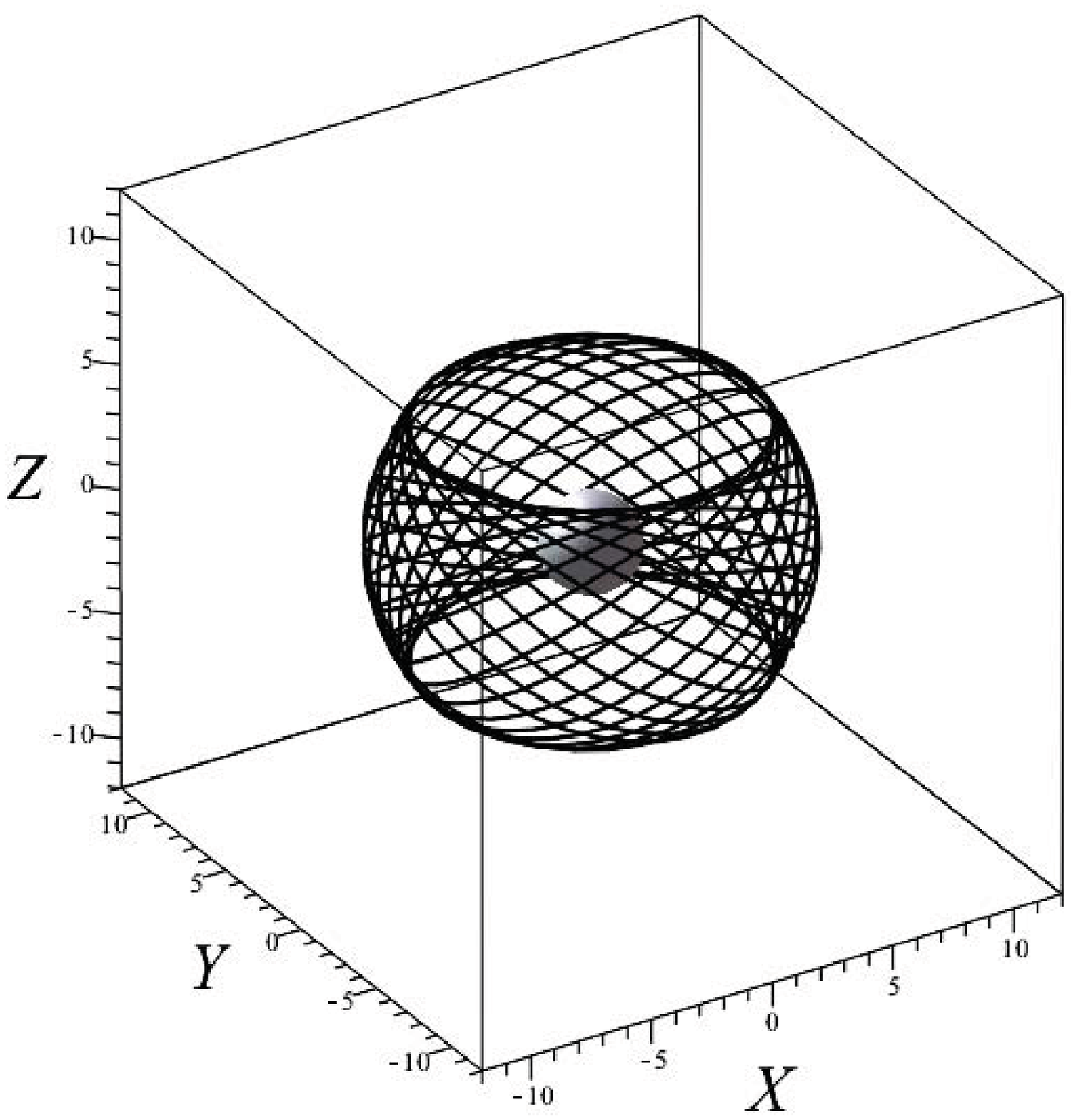}& 
\includegraphics[scale=0.35]{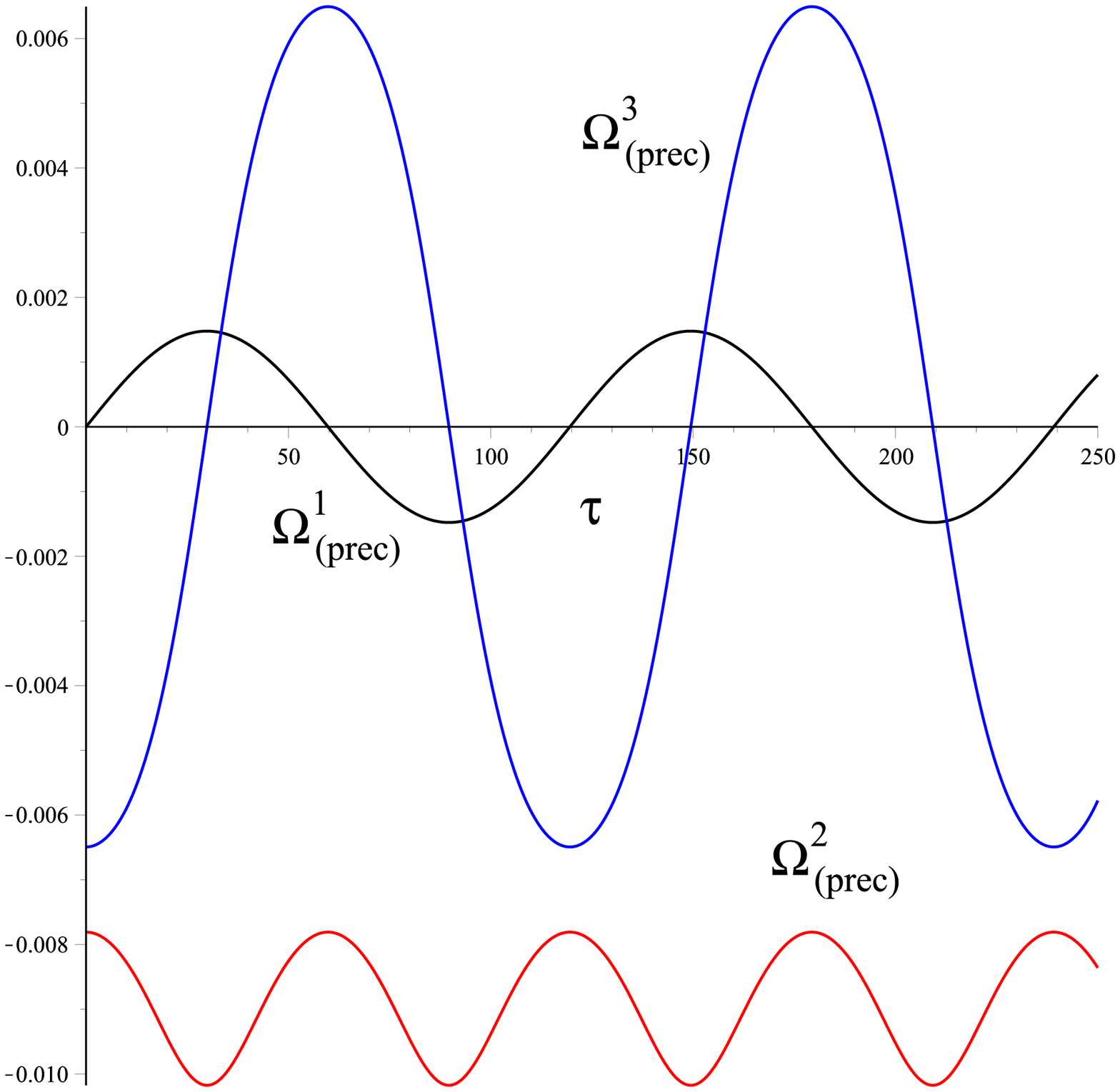}\cr
(a) & (b)
\end{array}
\]
\caption{An example of a spherical orbit at $r=r_0$ is shown in panel (a) for the choice of parameters $M=1$, $a/M=0.5$, $r_0/M=8$ and $K/M^2=9$, implying that $E\approx0.9446$, $L/M\approx2.6880$ and $\theta_-\approx0.9266$.
The inclination angle is $\iota\approx36.9583$ degrees.
Initial conditions are chosen to be $\theta(0)=\pi/2$ and $\phi(0)=0$ with $\epsilon_\theta=-1$, so that $\dot \theta(0)\approx-0.03160$. 
The corresponding evolution of the components of the gyroscope precession frequency along the orbit are shown in panel (b).
}
\label{fig:3}
\end{figure*}

\begin{figure*}
\[
\begin{array}{cc}
\includegraphics[scale=0.35]{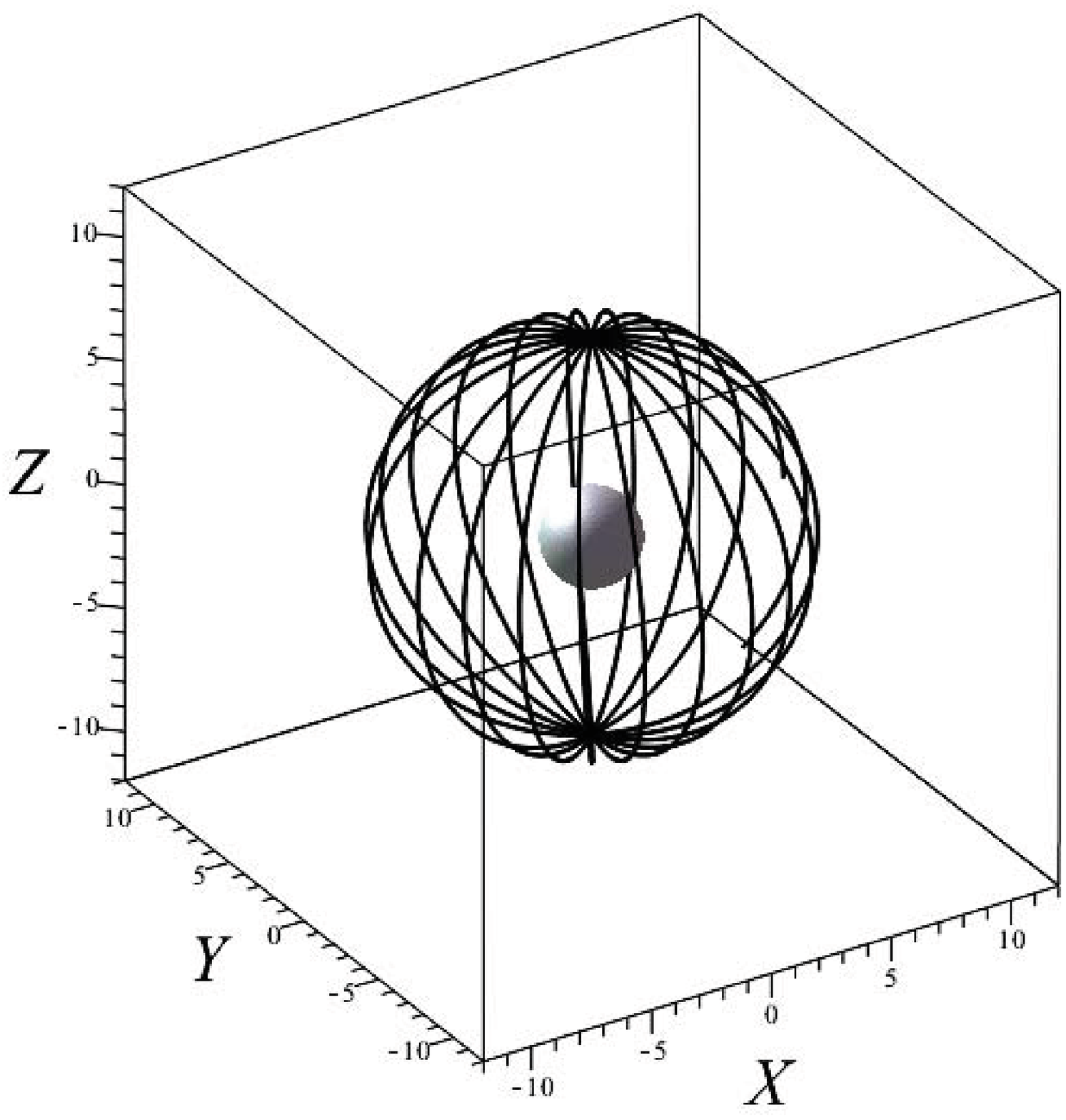}& 
\includegraphics[scale=0.35]{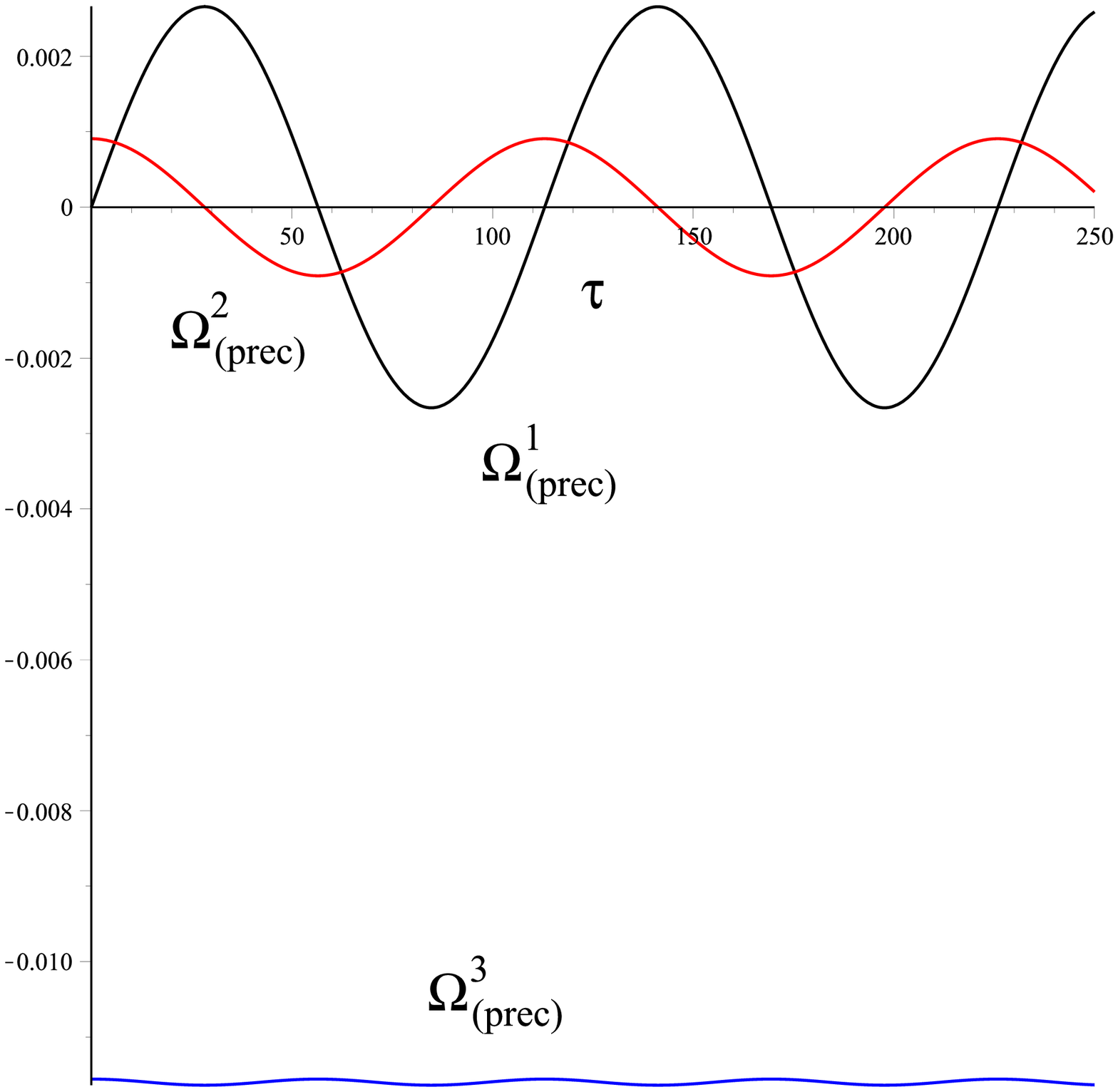}\cr
(a) & (b)
\end{array}
\]
\caption{An example of a polar orbit at $r=r_0$ is shown in panel (a) for the choice of parameters $M=1$, $a/M=0.5$ and $r_0/M=8$, implying that $E\approx0.9484$ and $K/M^2\approx12.9604$.
Initial conditions are chosen to be $\theta(0)=\pi/2$ and $\phi(0)=0$ with $\epsilon_\theta=-1$, so that $\dot \theta(0)\approx-0.05576$. 
The polar angle $\theta$ is allowed to decrease here without limit along the orbit since $\dot \theta$ never changes its sign, identifying values outside the usual range $[0,\pi]$ in the obvious way.
The corresponding evolution of the components of the gyroscope precession angular velocity along the orbit are shown in panel (b).
}
\label{fig:4}
\end{figure*}

\begin{figure*}
\[
\begin{array}{cc}
\includegraphics[scale=0.35]{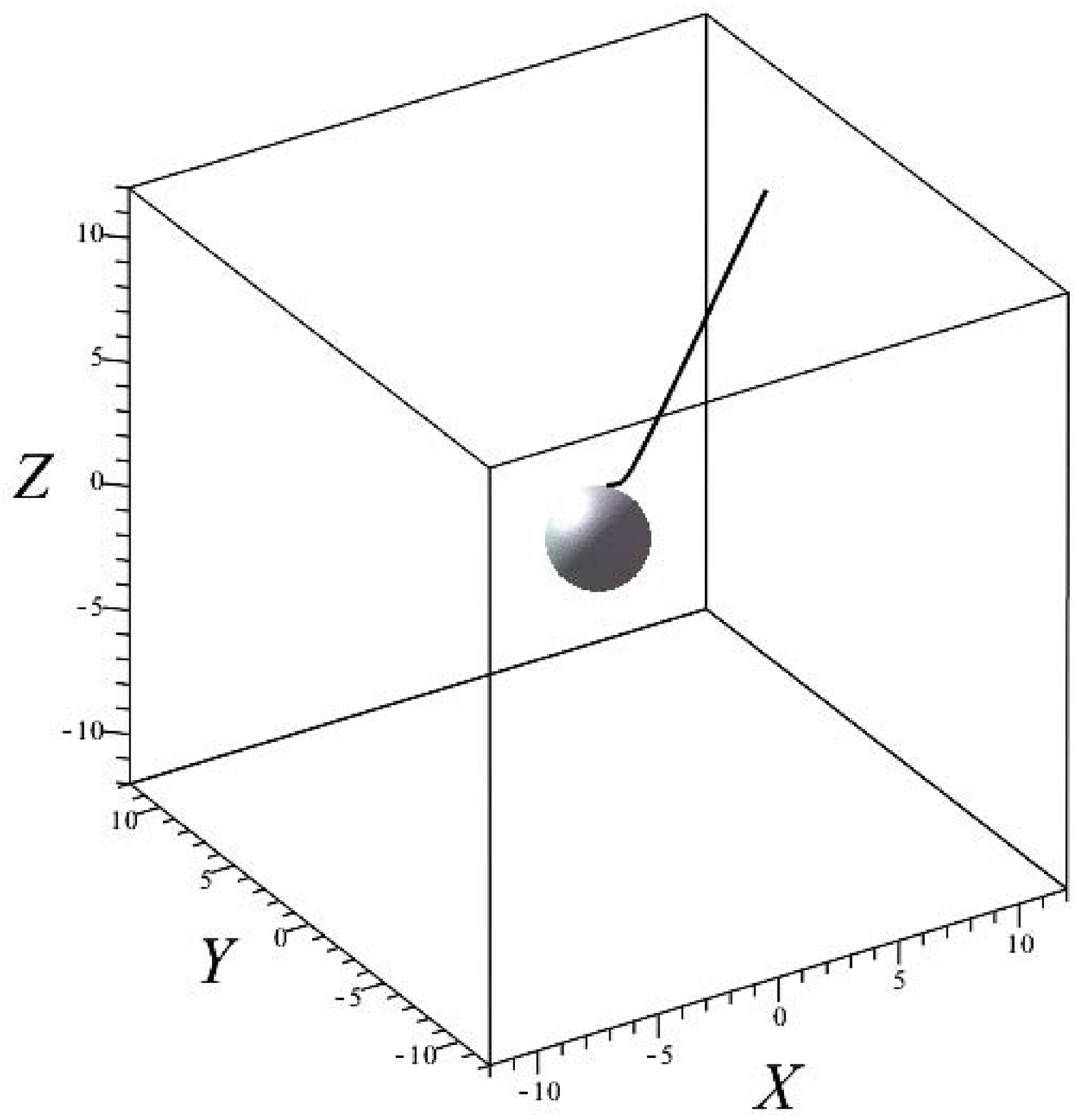}& 
\includegraphics[scale=0.35]{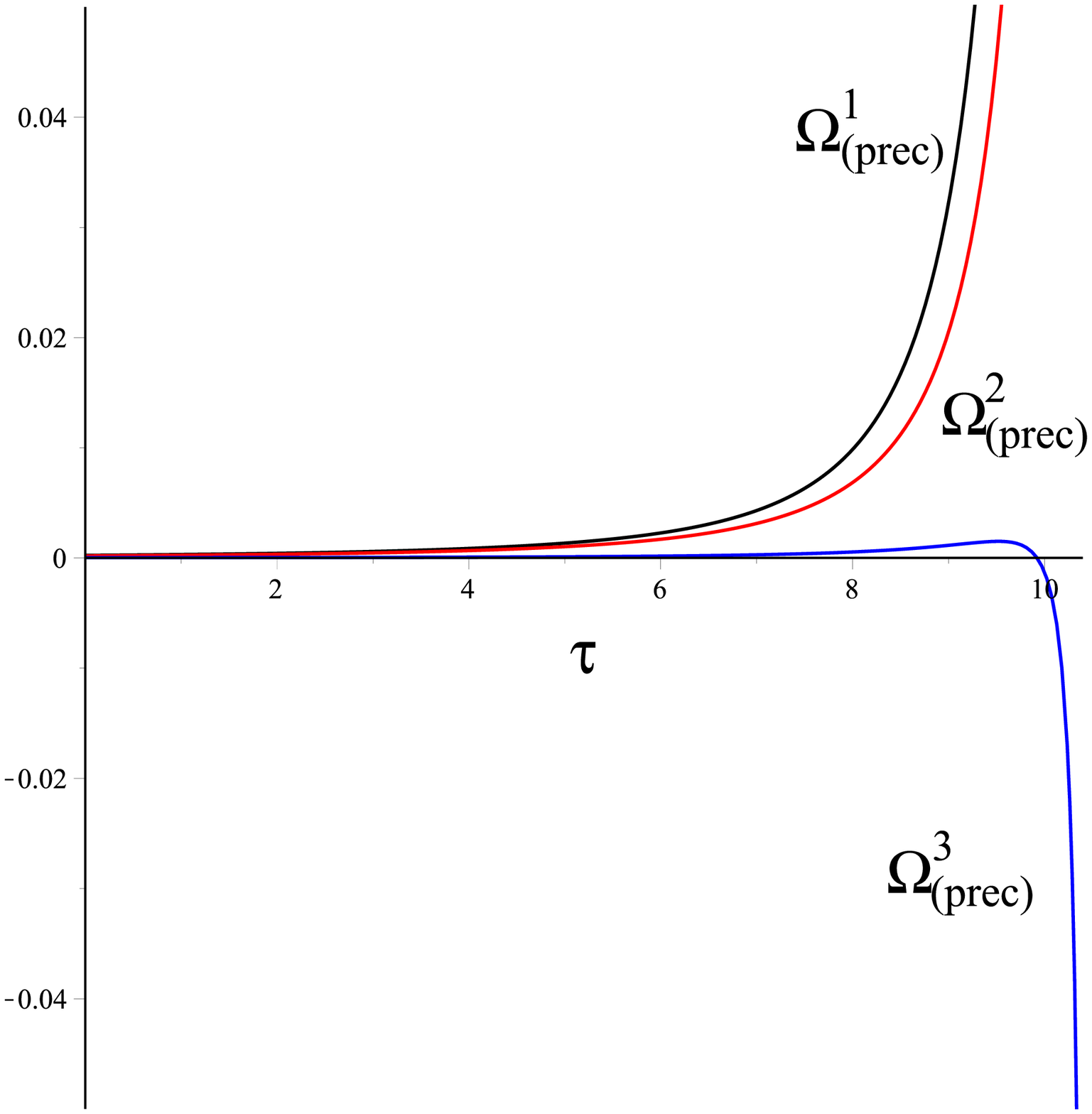}\cr
(a) & (b)
\end{array}
\]
\caption{An example of an orbit with fixed $\theta=\theta_0$ is shown in panel (a) for the choice of parameters $M=1$, $a/M=0.5$, $\theta_0=\pi/6$ and $K/M^2=1$, implying that $E\approx1.9415$ and $L/M\approx-0.2080$.
Initial conditions are chosen to be $r(0)/M=20$ and $\phi(0)=0$ with $\epsilon_r=-1$, so that $\dot r(0)\approx-1.6939$. 
The integration stops at the ergosphere.
The corresponding evolution of the components of the gyroscope precession angular velocity along the orbit is shown in panel (b).
}
\label{fig:5}
\end{figure*}

Typical examples of precession along general bound and unbound geodesic orbits are shown in Figs.~\ref{fig:1} and \ref{fig:2}, respectively. It is important to keep in mind for interpretational purposes that modulo a boost, the angular velocity component indices 1,2,3 are aligned with the $r,\theta,\phi$ directions.
Special orbits are considered too (see Figs.~\ref{fig:3}--\ref{fig:5}). The Cartesian axes shown in the plots are associated with the naive Cartesian coordinates related to the Boyer-Lindquist spherical coordinates by the same definitions as in flat space, namely \eqref{xyz}. The black hole outer horizon is shown as a gray sphere at the origin.
As a general feature we see that the frame components of the precession angular velocity have oscillating behavior with varying amplitude (Figs.~\ref{fig:1}--\ref{fig:2}), except for the case of spherical orbits (Figs.~\ref{fig:3}--\ref{fig:4}), where the amplitude is mainly constant. 
At the ergosphere where the static observers have their horizon the $|{\Omega}_{\rm(prec)}^i|$ increase due to the diverging of the overall $\gamma$ factor in Eqs.~\eqref{Omega} (see Fig.~\ref{fig:5}).

If instead of $E(U,m)_a$ one uses the Frenet-Serret frame $\{e_a\}$ related by
\beq\label{relrotationmarckboosted}
e_a
=E(U,m) \,\R^b{}_a 
\,,
\eeq
the following relation holds 
\beq\label{DEUm}
\frac{D E(U,m)_a}{d\tau} 
= \frac{D e_b}{d\tau} [\R^{-1}]^b{}_a 
+ E(U,m)_b [\R\,\dot\R{}^{-1}]^b{}_a\,,
\eeq
so that introducing 
\beq
 [\R\,\dot\R{}^{-1}]^b{}_a =\epsilon_{bca} \Omega_\R{}^c\,,\
  \Omega_\R = \Omega_\R{}^a E(U,m)_a
\eeq
and using  Eq.~\eqref{omegaT} for the first term in \eqref{DEUm}, one has
\beq
\Omega(U,m)\times_U E(U,m)_a
= [ \Omega_{\rm(FS)}+ \Omega_\R{}]\times_U  E(U,m)_a\,. 
\eeq
 
\begin{widetext}

Tedious direct evaluation of this rotation matrix following the sequence of boosts and rotations described above in the construction of the Frenet-Serret frame yields the result
\begin{eqnarray}
\R^1{}_1
&=&
-\frac{\gamma\Sigma^{1/2}}{(1+\gamma)E(K\Delta)^{1/2}}\left\{
a^2\cos\theta\sin\theta\coth\beta\,\dot r \dot \theta
-\frac{r\tanh\beta}{\Sigma^2}\left[
\Delta\Sigma\cosh^2\beta
+P\left(aB+\frac{E\Sigma}{\gamma}\right)
\right]
\right\}
\,,\nonumber\\
\R^2{}_1
&=&
-\frac{\gamma\Sigma^{1/2}}{(1+\gamma)EK^{1/2}}\left\{
r\tanh\beta\,\dot r \dot \theta
+\frac{a\cot\theta\coth\beta}{\Sigma^2}\left[
a\Sigma\sin^2\theta\sinh^2\beta
+B\left(P+\frac{E\Sigma}{\gamma}\right)
\right]
\right\}
\,,\nonumber\\
\R^3{}_1
&=&
\frac{\gamma}{(1+\gamma)E(\Delta K)^{1/2}}\left[
\frac{r\tanh\beta}{\sin\theta}\dot r\left(a\sin^2\theta-\frac{EB}{\gamma}\right)
+a\cos\theta\coth\beta\,\dot \theta\left(\Delta+\frac{EP}{\gamma}\right)
\right]
\,,\nonumber\\
\R^1{}_2
&=&
\frac{a\gamma\Sigma^{1/2}}{(1+\gamma)E(K\Delta)^{1/2}}\left\{
r\sin\theta \,\dot r \dot \theta
+\frac{\cos\theta}{\Sigma^2}\left[
\Delta\Sigma\cosh^2\beta
+P\left(aB+\frac{E\Sigma}{\gamma}\right)
\right]
\right\}
\,,\nonumber\\
\R^2{}_2
&=&
-\frac{\gamma\Sigma^{1/2}}{(1+\gamma)EK^{1/2}}\left\{
a\cos\theta \,\dot r \dot \theta
-\frac{r}{\Sigma^2\sin\theta}\left[
a\Sigma\sin^2\theta\sinh^2\beta
+B\left(P+\frac{E\Sigma}{\gamma}\right)
\right]
\right\}
\,,\nonumber\\
\R^3{}_2
&=&
\frac{\gamma}{(1+\gamma)E(\Delta K)^{1/2}}\left[
a\cot\theta\,\dot r\left(a\sin^2\theta-\frac{EB}{\gamma}\right)
-r\,\dot \theta\left(\Delta+\frac{EP}{\gamma}\right)
\right]
\,,\nonumber\\
\R^1{}_3
&=&
-\frac{\gamma\,\dot r}{(1+\gamma)E(\Delta\Sigma)^{1/2}\cosh\beta\sinh\beta}\left(
aB-\frac{E\Sigma}{\gamma}\sinh^2\beta
\right)
\,,\nonumber\\
\R^2{}_3
&=&
\frac{\gamma\,\dot \theta}{(1+\gamma)E\Sigma^{1/2}\cosh\beta\sinh\beta}\left(
P+\frac{E\Sigma}{\gamma}\cosh^2\beta
\right)
\,,\nonumber\\
\R^3{}_3
&=&
\frac{\gamma}{(1+\gamma)E\Sigma\Delta^{1/2}\sin\theta\cosh\beta\sinh\beta}\left[
\left(\Delta+\frac{EP}{\gamma}\right)B
-\sinh^2\beta\left(2MrB-\Sigma L\right)
\right]
\,.
\end{eqnarray}
The latter components are evaluated using the geodesic equations to replace second proper time derivatives in $ \dot\R{}^{-1}$ in terms of first time derivatives and constants of the motion.

For motion confined to the equatorial plane, this rotation matrix reduces to a rotation in the $r$-$\phi$ plane of the tangent space
\beq
(\R^b{}_a)
= \begin{pmatrix}
\cos \Lambda & 0 & \sin \Lambda \cr
0 & 1 & 0 \cr
-\sin \Lambda & 0 & \cos\Lambda \cr
\end{pmatrix}
\,,
\eeq
with
\footnote{
Note that the expression for $\cos\Lambda$ in Eq.~\eqref{cosLambdadef} should reduce to the corresponding equation (C16) of Ref.~\cite{Bini:2016iym}, but the latter is misprinted with the term $-(a-Nx^2)^2$ instead of $(a-Nx)^2$. 
}

\begin{eqnarray}
\label{cosLambdadef}
\cos \Lambda&=&\frac{\gamma}{(1+\gamma)Er^2\Delta^{1/2}(r^2+x^2)^{1/2}}\left[
\Delta(r^2+x^2)-(ax-r^2E)\left(ax+\frac{Er^2}{\gamma}\right)\right]\nonumber\\
&=&
\frac1{\sqrt{\Delta(r^2+x^2)}}\left[r^2N+ax+\frac{(a-Nx)^2}{E+N}\right]
\,,\nonumber\\
\sin \Lambda&=&\frac{\gamma r\,\dot r}{(1+\gamma)E\Delta^{1/2}(r^2+x^2)^{1/2}}
\left(a-\frac{Ex}{\gamma}\right)
\,,
\end{eqnarray}
where $N=\sqrt{1-2M/r}$.

\end{widetext}
In this special case, the only nonvanishing component \eqref{Omegab2equat} of the associated angular velocity can then be written as
\beq
\label{Omegab2equatn}
\Omega(U,m)^2\vert_{\theta=\pi/2}=\frac{d\Lambda}{d\tau}-{\mathcal T}\,,
\eeq
with ${\mathcal T}=(a+Ex)/(r^2+x^2)$.
Finally, the precession angular velocity is given by Eq.~\eqref{Omegaprec}, which in the case of equatorial plane motion (where $E(U,m)_2=e_2=e_{\hat\theta}$) reduces to 
\beq
\label{omega_prec_equat}
{\Omega}_{\rm(prec)}\vert_{\theta=\pi/2}
=-\left(\frac{L-2Mx/r}{\Delta}-\frac{a+Ex}{r^2+x^2}+\frac{d\Lambda}{d\tau}\right)e_2\,.
\eeq
The last term in this expression corresponds to the angular velocity of the Wigner rotation discussed in Appendix C of Ref.~\cite{Bini:2016iym}.
Note the opposite sign when comparing Eq.~\eqref{omega_prec_equat} with the analogous equations in Ref.~\cite{Bini:2016iym} (Eqs.~(41) and (44)) and in Ref.~\cite{Bini:2016ovy} (Eqs.~(48) and (50)), which refer to the counterclockwise angular velocity in the plane, corresponding to the component of this vector along the upward direction orthogonal to the equatorial plane $e_z=-e_{\hat\theta}=-e_2$.

\section{Concluding remarks}

Generalizing our understanding of the simpler case of planar geodesic motion,
we have evaluated the precession angular velocity of the parallel transported spin vector of a gyroscope which moves along a general timelike geodesic in the Kerr spacetime. The precession is measured with respect to the celestial sky at spatial infinity represented locally by the static observer Boyer-Lindquist coordinate grid. 

We have defined various frames in the local rest space along the gyro world line by boosting natural frames adapted to special families of observers linked to the symmetries of the Kerr spacetime. In particular, we have considered the usual static (distantly nonrotating) observers and the ZAMOs (locally nonrotating observers), and the boosts of their natural spherical frames, as well as the Frenet-Serret frame used in the construction of Marck's parallely transported frame  along a general timelike geodesic world line.

We have explicitly computed the components of the gyroscope precession with respect a frame in the local rest space of the gyro world line which represents axes locked to the distant fixed Cartesian directions, removing the aberration of those directions due to the relative motion of the gyro and the static grid.
We then illustrated these general results by discussing several examples of both general bound and unbound nonequatorial plane orbits as well as special cases of orbits at constant radius (spherical and polar) and those with constant polar angle $\theta$.

By allowing the initial observer family to be unspecified in our discussion in the genuine spirit of relativity, we were able to perform an equivalent evaluation of the spin-precession angular velocity for the ZAMO frame. This leads to a geometrical interpretation of the various contributions to the spin-orbit Hamiltonian derived in Ref.~\cite{Barausse:2009aa}.
We have also clarified the construction of Marck's parallely transported frame along a timelike geodesic in terms of the Frenet-Serret frame obtained by two successive boosts of the natural spherical frame from the local rest space of the Carter observers, which are key to the separation of the geodesic equations of motion. This Frenet-Serret frame played a simplifying role in the analysis of the special case of bound and unbound timelike geodesic orbits confined to the equatorial plane, the natural extension of which is the present article.  General timelike geodesic motion significantly complicates matters, and only an approach which uses well defined geometrical objects (transport laws, projections, boosts, etc.) can  bring a clearer understanding of the physical properties underlying spin precession.

\subsection*{Acknowledgments}
D.B. thanks the Italian INFN (Naples) for partial support.
All the authors are grateful to the International Center for Relativistic Astrophysics Network based in Pescara, Italy for partial support.

\appendix

\section{Static and ZAMO observer families}

\vspace{1cm}

\subsection{Kinematical properties of static observers}

The static observers are accelerated, with 4-acceleration
\begin{eqnarray}
a(m)&=&\frac{M\sqrt{\Delta}(r^2-a^2\cos^2\theta)}{\Sigma^{3/2}(\Delta -a^2 \sin^2\theta)}e_{\hat r}\nonumber\\
&& -
\frac{2Mra^2 \sin \theta \cos \theta}{\Sigma^{3/2}(\Delta -a^2 \sin^2\theta)}e_{\hat \theta}\,,
\end{eqnarray}
and are locally rotating, with vorticity vector
\begin{eqnarray}
\omega(m)&=&-\frac{2aMr \sqrt{\Delta}\cos\theta}{\Sigma^{3/2}(\Delta-a^2 \sin^2\theta)} e_{\hat r}\nonumber\\
&& -\frac{Ma(r^2-a^2\cos^2\theta)\sin \theta}{\Sigma^{3/2}(\Delta -a^2\sin^2\theta)}e_{\hat \theta}\,,
\end{eqnarray}
but the  congruence of their world lines is not expanding, i.e., has vanishing expansion
$\theta(m)=0$, due to the alignment of their 4-velocity with the timelike Killing direction (see, e.g., Ref.~\cite{Jantzen:1992rg} for a detailed description of the kinematical properties of a congruence of world lines, including acceleration, vorticity and expansion). 

For completeness we review properties of the static observer adapted frame including the transport laws along the geodesic congruence $U$ decomposed as in Eqs.~\eqref{Uthd}--\eqref{Uthd2}.
The components of the spatial gravitational force, the Fermi-Walker and spatial curvature rotation vectors
(see Eqs.~\eqref{delUE}--\eqref{omgUu} with $u=m$) are given by

\begin{widetext}

\begin{eqnarray}
F^{(G)}_{({\rm fw},U,m)}{}^1&=&
\gamma(U,m)\frac{M(r^2-a^2\cos^2\theta)}{\Sigma^{3/2}(\Sigma-2Mr)}[a\sin\theta\,\nu(U,m)^3-\sqrt{\Delta}]
\,,\nonumber\\
F^{(G)}_{({\rm fw},U,m)}{}^2&=&
\gamma(U,m)\frac{2aMr\cos\theta}{\Sigma^{3/2}(\Sigma-2Mr)}[a\sin\theta-\sqrt{\Delta}\,\nu(U,m)^3]
\,,\nonumber\\
F^{(G)}_{({\rm fw},U,m)}{}^3&=&
\gamma(U,m)\frac{aM}{\Sigma^{3/2}(\Sigma-2Mr)}[-(r^2-a^2\cos^2\theta)\sin\theta\,\nu(U,m)^1+2r\sqrt{\Delta}\cos\theta\,\nu(U,m)^2]
\,.
\end{eqnarray}
Moreover
\beq
\omega_{({\rm fw},m)}{}^1=\frac{2aMr\cos\theta\sqrt{\Delta}}{\Sigma^{3/2}(\Sigma-2Mr)}\,,\qquad
\omega_{({\rm fw},m)}{}^2=\frac{aM\sin\theta(r^2-a^2\cos^2\theta)}{\Sigma^{3/2}(\Sigma-2Mr)}\,,
\eeq
and
\begin{eqnarray}
\omega_{({\rm sc},U,m)}{}^1&=&
-\frac{\cos\theta\nu(U,m)^3}{\Sigma^{3/2}(\Sigma-2Mr)\sin\theta}[\Sigma^2-4Mr\Sigma+2Mr(r^2+a^2)]
+\frac{4aMr\cos\theta\sqrt{\Delta}}{\Sigma^{3/2}(\Sigma-2Mr)}
\,,\nonumber\\
\omega_{({\rm sc},U,m)}{}^2&=&
\frac{\nu(U,m)^3}{\sqrt{\Delta}\Sigma^{3/2}(\Sigma-2Mr)}[(r-M)\Sigma^2-M(r^2-a^2)\Sigma-2Mr^2\Delta]
\,,\nonumber\\
\omega_{({\rm sc},U,m)}{}^3&=&-\frac{1}{\Sigma^{3/2}}[a^2\sin\theta\cos\theta\nu(U,m)^1+r\sqrt{\Delta}\nu(U,m)^2]
\,.
\end{eqnarray}

\end{widetext}

\subsection{Kinematical properties of ZAMOs}

We record the key properties of ZAMOs whose 4-velocity is orthogonal to the Boyer-Lindquist time coordinate hypersurfaces
\beq
\label{n}
n\equiv e_{\hat t}=N^{-1}(\partial_t-N^{\phi}\partial_\phi)\,,
\eeq
with corresponding 1-form
\beq
n^\flat =-\omega^{{\hat t}}=-N d t\,,
\eeq
where $N=(-g^{tt})^{-1/2}$ and $N^{\phi}=g_{t\phi}/g_{\phi\phi}$ are the lapse and shift functions, respectively. 

The accelerated ZAMOs are locally nonrotating in the sense that their vorticity vector $\omega(n)$ vanishes, but they have a nonzero expansion tensor $\theta(n)$ with vanishing expansion scalar $\theta(n)^\alpha{}_\alpha$ so that it agrees with the shear tensor. Its nonzero components 
can be described by a shear vector $\theta_{\hat \phi}(n)^\alpha$
\begin{eqnarray}
\label{exp_zamo}
\theta(n) &=& e_{\hat\phi}\otimes\theta_{\hat\phi}(n)
           +\theta_{\hat\phi}(n)\otimes e_{\hat\phi}\nonumber\\
\theta_{\hat \phi}(n)^\alpha &=& \theta(n)^{\hat\beta}{}_{\hat\phi}\,e_{\hat\beta}{}^\alpha
\,.
\end{eqnarray} 

The nonzero ZAMO kinematical quantities (acceleration $a(n)=\nabla_n n$ and shear tensor) as well as the curvature vectors associated with the diagonal metric coefficients \cite{Jantzen:1992rg,Bini:1997ea,Bini:1997eb,Bini:1999wn} only have nonzero components in the $r$-$\theta$ 2-plane of the tangent space, i.e.,
\begin{eqnarray}
\label{accexp}
a(n) & = & a(n)^{\hat r} e_{\hat r} + a(n)^{\hat\theta} e_{\hat\theta}\nonumber\\
& =& \partial_{\hat r}(\ln N) e_{\hat r} + \partial_{\hat\theta}(\ln N)  e_{\hat\theta}
\,,
\nonumber\\
\theta_{\hat\phi}(n) 
& = & \theta(n)^{\hat r}{}_{\hat\phi} e_{\hat r} + \theta(n)^{\hat\theta}{} _{\hat\phi}e_{\hat \theta} \nonumber\\
& = & -\frac{\sqrt{g_{\phi\phi}}}{2N}\,(\partial_{\hat r} N^\phi e_{\hat r} + \partial_{\hat\theta} N^\phi e_{\hat \theta})
\,, 
\nonumber\\
\kappa(x^i,n)
& = & \kappa(x^i,n)^{\hat r} e_{\hat r} + \kappa(x^i,n)^{\hat\theta} e_{\hat\theta}\nonumber\\
& = & -[\partial_{\hat r}(\ln \sqrt{g_{ii}}) e_{\hat r} + \partial_{\hat\theta}(\ln \sqrt{g_{ii}})e_{\hat\theta}]
\,.
\end{eqnarray}
We have then three $\kappa(x^i,n)$ ``coordinate" curvature vectors: $\kappa(r,n)^i$, $\kappa(\theta,n)^i$ and $\kappa(\phi,n)^i$, all belonging to the $\hat r$-$\hat \theta$ plane.  
In the static limit (as it is the case of a Schwarzschild black hole) $N^\phi\to0$ and the expansion vector $\theta_{\hat\phi}(n)$ vanishes. 
We list below the nonvanishing components of the kinematical fields:
acceleration
\begin{eqnarray}
a(n)^{\hat r}&=&-\frac{M}{\sqrt{\Delta}\Sigma^{3/2}A}
\left\{a^2\cos^2\theta[(r^2+a^2)^2-4Mr^3]\right.\nonumber \\
&&\left.-r^2[(r^2+a^2)^2-4a^2Mr]\right\}\,,
\nonumber \\
a(n)^{\hat \theta}&=&-\frac{ 2\sin\theta\cos\theta Mr a^2 (r^2+a^2)}{\Sigma^{3/2}A} 
\,,
\end{eqnarray}
with $A=(r^2+a^2)^2-a^2\Delta\sin^2\theta$,
shear tensor
\begin{eqnarray}
\theta(n)^{\hat r \hat \phi}&=& -\frac{aM\sin\theta}{\Sigma^{3/2}A}[r^2(3r^2+a^2)\nonumber\\
&&
+a^2(r^2-a^2)\cos^2\theta)]
\,,\nonumber \\
\theta(n)^{\hat \theta \hat \phi}&=&\frac{2ra^3M\sin^2\theta\cos\theta\sqrt{\Delta}}{\Sigma^{3/2}A}\,,
\end{eqnarray}
and curvature vectors
\begin{eqnarray}
\kappa(r,n)^{\hat r}&=& -\frac{r\Delta-(r-M)\Sigma}{\Sigma^{3/2}\sqrt{\Delta}}
\,,\nonumber \\
\kappa(r,n)^{\hat\theta}&=& \frac{a^2\sin\theta\cos\theta}{\Sigma^{3/2}}
\,,\nonumber \\
\kappa(\theta,n)^{\hat r}&=& -\frac{r\sqrt{\Delta}}{\Sigma^{3/2}}
\,,\nonumber \\
\kappa(\theta,n)^{\hat\theta}&=& \kappa(r,n)^{\hat\theta} 
\,,\nonumber \\
\kappa(\phi,n)^{\hat r}&=& -\frac{r\Sigma^2-Ma^2\sin^2\theta(r^2-a^2\cos^2\theta)}{\Sigma^{3/2}A}\sqrt{\Delta}
\,,\nonumber \\
\kappa(\phi,n)^{\hat\theta}&=& -\frac{(r^2+a^2)A-a^2\sin^2\theta\Delta\Sigma}{\Sigma^{3/2}A\sin\theta}\cos\theta
\,.
\end{eqnarray}

The nonvanishing components of the shear vector enter the transport law for the spatial triad $e_{\hat a}$ along the world line of $n$, i.e., 
\begin{eqnarray}
P(n)\nabla_{\hbox{$n$}} e_{\hat r}&=&\omega_{({\rm fw},n)}{}^{\hat \theta}e_{\hat \phi}\,,\nonumber\\  
P(n)\nabla_{\hbox{$n$}} e_{\hat \theta}&=&-\omega_{({\rm fw},n)}{}^{\hat r}e_{\hat \phi}\,, \nonumber\\ 
P(n)\nabla_{\hbox{$n$}} e_{\hat \phi}&=&-\omega_{({\rm fw},n)}{}^{\hat \theta}e_{\hat r}+\omega_{({\rm fw},n)}{}^{\hat r}e_{\hat \theta}\,,
\end{eqnarray}
where $\omega_{({\rm fw},n)}$ is the Fermi-Walker angular velocity with components
\beq
\omega_{({\rm fw},n)}{}^{\hat r}=\theta(n)^{\hat \theta \hat \phi}\,,\qquad
\omega_{({\rm fw},n)}{}^{\hat \theta}=-\theta(n)^{\hat r \hat \phi}\,.
\eeq
In terms of the cross product, we have
\beq
P(n)\nabla_{\hbox{$n$}} e_{\hat a} =-\omega_{({\rm fw},n)} \times_n e_{\hat a} \,.
\eeq

A timelike test particle's 4-velocity $U$ can be decomposed with respect to the ZAMOs as in Eq.~\eqref{Uzamo}, i.e., $U=\gamma(U,n)(n + \nu(U,n))$.
Evaluating the derivative along $U$ of the ZAMO adapted frame \eqref{zamotriad} leads to  
\beq
P(n)\nabla_U e_{\hat a}= -\gamma(U,n) [\omega_{({\rm fw},n)}+\omega_{({\rm sc},U,n)}]\times_n e_{\hat a}\,,
\eeq
with
\begin{eqnarray}
\omega_{({\rm sc},U,n)}{}^{\hat r}  &=&  \nu(U,n)^{\hat \phi}\kappa(\phi,n)^{\hat \theta}\,, \nonumber\\
\omega_{({\rm sc},U,n)}{}^{\hat \theta}&=& -\nu(U,n)^{\hat \phi}\kappa(\phi,n)^{\hat r}\,,\nonumber\\
\omega_{({\rm sc},U,n)}{}^{\hat \phi}&=& -[\nu(U,n)^{\hat r}\kappa(r,n)^{\hat \theta}-\nu(U,n)^{\hat \theta}\kappa(\theta,n)^{\hat r}]\,. \nonumber\\
\end{eqnarray}
Finally, the Fermi-Walker spatial gravitational force defining the geodesic precession angular velocity 
(see Eqs.~\eqref{delUE}--\eqref{omgUu} with $u=n$) is given by 
\beq
F^{(G)}_{({\rm fw},U,n)}=-\nabla_Un
=-\gamma(U,n)[a(n)+\theta(n)\rightcontract \nu(U,n)]\,.
\eeq

\section{Marck's frame, tidal  matrices and diagonalization properties}

The nonvanishing components of the electric part of the Riemann tensor $E(U)_{\alpha\beta}=R_{\alpha\mu\beta\nu}U^\mu U^\nu$ in the parallel propagated frame $\{E_i\}$ computed explicitly in Ref.~\cite{Bini:2011zzc} are given by 

\begin{widetext}

\begin{eqnarray}
\label{eleriem_par}
E(U)_{ 1 1}&=&-\frac{3Mr}{\Sigma^3K}J_3\sinh^2\beta\cosh^2\beta\cos^2\Psi+\frac{Mr}{\Sigma^3}J_5\,, \nonumber\\
E(U)_{ 1 2}&=&-\frac{3Ma\cos\theta}{\Sigma^3K}\sinh \beta\cosh \beta(J_1\cosh^2\beta-4r^2J_4)\cos\Psi\,, \nonumber\\
E(U)_{ 1 3}&=&-\frac{3Mr}{\Sigma^3K}J_3\sinh^2\beta\cosh^2\beta\cos\Psi\sin\Psi\,, \nonumber\\
E(U)_{ 2 2}&=&\frac{Mr}{\Sigma^3K}[3J_3\cosh^4\beta-\cosh^2\beta(J_1-8a^2\cos^2\theta J_2)+2r^2J_5]\,, \nonumber\\
E(U)_{ 2 3}&=&-\frac{3Ma\cos\theta}{\Sigma^3K}\sinh \beta\cosh \beta(J_1\cosh^2\beta-4r^2J_4)\sin\Psi\,, \nonumber\\
E(U)_{ 3 3}&=&\frac{3Mr}{\Sigma^3K}J_3\sinh^2\beta\cosh^2\beta\cos^2\Psi-\frac{Mr}{\Sigma^3K}[3J_3\cosh^4\beta\nonumber\\
&&-4\cosh^2\beta(J_3+2a^2\cos^2\theta J_4)+r^2J_5]\,, \nonumber\\
\end{eqnarray}
where
\begin{eqnarray}
J_1&=&5r^4-10r^2a^2\cos^2\theta+a^4\cos^4\theta\,, \nonumber\\
J_2&=&3r^2-a^2\cos^2\theta\,, \nonumber\\
J_3&=&r^4-10r^2a^2\cos^2\theta+5a^4\cos^4\theta=J_1-4J_4\,, \nonumber\\
J_4&=&r^2-a^2\cos^2\theta\,, \nonumber\\
J_5&=&r^2-3a^2\cos^2\theta\,.
\end{eqnarray}

The nonvanishing components of the magnetic part of the Riemann tensor $H(U)_{\alpha\beta}={}^*R_{\alpha\mu\beta\nu}U^\mu U^\nu$ in the parallel propagated frame are 
\begin{eqnarray}
H(U)_{ 1  1}&=&\frac{aM\cos\theta}{\Sigma^3}\left(J_2-\frac{3J_1}{K}\sinh^2\beta\cosh^2\beta\cos^2\Psi\right)\,, \nonumber\\
H(U)_{ 1  2}&=&\frac{3Mr}{\Sigma^4}\sinh \beta\cosh \beta\cos\Psi\left(J_3-\frac{a^2\cos^2\theta}{K}J_1\right)\,, \nonumber\\
H(U)_{ 1  3}&=&-\frac{3aM\cos\theta}{\Sigma^3K}\sinh^2\beta\cosh^2\beta\sin\Psi\cos\Psi J_1\,, \nonumber\\
H(U)_{ 2  2}&=&\frac{3aM\cos\theta}{\Sigma^5K}\left\{(K^2-r^2a^2\cos^2\theta)J_1+\frac{K}{3}
[J_1J_2-r^2(J_1-3J_3+4\Sigma J_4)]
\right\}\,, \nonumber\\
H(U)_{  2  3}&=&\frac{3Mr}{\Sigma^4}\sinh \beta\cosh \beta\sin\Psi\left(J_3-\frac{a^2\cos^2\theta}{K}J_1\right)\,, \nonumber\\
H(U)_{  3  3}&=&\frac{3aM\cos\theta}{\Sigma^3K}\left\{\left(\sinh^2\beta\cosh^2\beta\cos^2\Psi-\frac{K^2-r^2a^2\cos^2\theta}{\Sigma^2}\right)J_1\right.\nonumber\\
&&\left.-\frac{2K}{3\Sigma^2}
[J_1J_2-r^2(J_1+4\Sigma J_4)]
\right\}\,. 
\end{eqnarray}
\end{widetext}
Note that the sign of the magnetic part of the Riemann tensor depends on the sign convention chosen for the unit volume 4-form used to define the duality $*$ operation.

The components in the Frenet-Serret frame $\{e_i\}$ correspond to setting $\Psi=0$, for which both of these symmetric tensors reduce to block diagonal form with $e_3$ as an eigenvector. In that case the only off-diagonal component is  
\begin{eqnarray}
&&E(U)_{ 1 2}|_{\Psi=0}\\ \qquad&&=
-\frac{3Ma\cos\theta}{\Sigma^3K}\sinh \beta\cosh \beta(J_1\cosh^2\beta-4r^2J_4) \,.\nonumber
\end{eqnarray}
This vanishes on  the equatorial plane where $\cos\theta=0$, where the Marck frame diagonalizes the electric part of the curvature, while the magnetic part has the single nonvanishing component $H(U)_{ 1 2}$ there.
The eigenvector $e_3=e_1\times_U e_2$ is just the cross product of the direction of the angular part of the Carter frame relative velocity  with the parallel transported direction associated with the Killing form.

\end{document}